\documentclass[twocolumn,apsprd,showpacs, floatfix, 
amsmath,nofootinbib,preprintnumbers,letterpaper]{revtex4}

\usepackage{graphicx}
\usepackage{color}
\usepackage{verbatim}

\newcommand{\hMpc}{\,h^{-1}{\rm Mpc}}
\newcommand{\hMpcI}{\,h\,{\rm Mpc}^{-1}}
\newcommand{\vn}{\hat{n}}
\newcommand{\flux}{\mathcal{F}}

\def\lcdm{$\Lambda$CDM~}

\def\be{\begin{equation}}
\def\ee{\end{equation}}
\def\bea{\begin{eqnarray}}
\def\eea{\end{eqnarray}}
\newcommand{\vs}{\nonumber\\}
\newcommand{\Lya}{Lyman-$\alpha${ }}
\def\lg{\langle}
\def\rg{\rangle}
\usepackage{aas_macros}

\begin{document}

\preprint{FERMILAB-PUB-09-508-T}

\title{Cross-correlations of the Lyman-$\alpha$ forest with weak lensing convergence I:\\
Analytical Estimates of S/N and Implications for Neutrino Mass and Dark Energy}

 \author{Alberto Vallinotto}
 \email{avalli@fnal.gov}
 \affiliation{
 Center for Particle Astrophysics, Fermi National Accelerator Laboratory, 
 P.O. Box 500, Kirk Rd. \& Pine St., Batavia, IL 60510-0500 USA}

\author{Matteo Viel}
 \email{viel@oats.inaf.it}
 \affiliation{INAF - Osservatorio Astronomico di Trieste, Via G.B. Tiepolo 11,
I-34131 Trieste, Italy\\
INFN - National Institute for Nuclear Physics, Via Valerio 2,
I-34127 Trieste, Italy}

\author{Sudeep Das}
 \email{sudeep@astro.princeton.edu}
 \affiliation{Princeton University Observatory, Peyton Hall, Ivy Lane, Princeton, NJ 08544 USA\\
 Berkeley Center for Cosmological Physics, LBNL and Department of Physics, University of California, Berkeley, CA 94720.}

 \author{David N. Spergel}
 \email{dns@astro.princeton.edu}
 \affiliation{Princeton University Observatory,
Peyton Hall, Ivy Lane, Princeton, NJ 08544 USA}


\date{\today}

\begin{abstract}
We expect a detectable correlation between two seemingly unrelated quantities: the four point function of
the cosmic microwave background (CMB) and the amplitude of flux decrements in quasar (QSO) spectra.
The amplitude of CMB convergence in a given direction measures the projected surface density of matter.
Measurements of QSO flux decrements trace the small-scale distribution of gas along a given line-of-sight.  While
the cross-correlation between these two measurements is small for a single line-of-sight, upcoming large surveys should
enable its detection.  This paper presents analytical estimates for the signal to noise (S/N) for measurements
of the cross-correlation between the flux decrement and the convergence, $\langle\delta \flux \kappa\rangle$, and for measurements
of the cross-correlation between the variance in  flux decrement and the convergence, $\langle(\delta \flux)^2 \kappa\rangle$.  For the ongoing BOSS (SDSS III) and Planck surveys, we estimate an S/N of 30 and 9.6 for these two correlations.  For the proposed BigBOSS and ACTPOL surveys, we estimate an S/N of 130 and 50 respectively.  Since $\langle(\delta \flux)^2 \kappa\rangle \propto \sigma_8^4$, the amplitude of these cross-correlations can potentially be used to measure the amplitude of $\sigma_8$ at
$z \sim 2$ to 2.5\% with BOSS and Planck and even better with future data sets.  These measurements have the potential to test
alternative theories for dark energy and to constrain the mass of the neutrino.  The large potential signal estimated in our analytical calculations motivate tests with non-linear hydrodynamical simulations and analyses of upcoming data sets.
\end{abstract}

\maketitle

\section{Introduction}

The confluence of high resolution Cosmic Microwave Background (CMB) 
experiments and  large-scale spectroscopic surveys in the near future is 
expected to sharpen our view of the Universe. Arcminute 
scale CMB experiments such as Planck \cite{PLANCK}, the Atacama Cosmology Telescope \cite{ACT,hincks/etal:2009}, the 
South Pole Telescope  \cite{SPT,staniszewski/etal:2009}, QUIET  \cite{QUIET} and PolarBeaR \cite{POLAR}, will chart out the small scale 
anisotropies in the CMB.  This will shed new light on the primordial physics of inflation, 
as well 
as the astrophysics of the low redshift Universe through the signatures of the 
interactions of 
the CMB photons with large scale structure.   Spectroscopic surveys like BOSS \cite{mcdonald05,seljak05} and 
BigBOSS \cite{Schlegel:2009uw}
will trace the large scale structure of neutral gas, probing  the 
distribution and dynamics of matter in the Universe.  While these two datasets will be 
rich on 
their own, they will also complement and constrain each other. An interesting 
avenue for using the two datasets would be to utilize the fact that the arcminute-scale 
secondary 
anisotropies in the CMB are signatures of the same large scale structure that is  traced 
by 
the spectroscopic surveys, and study them in cross-correlation with each other. In this 
paper, we present the  
analytic estimates for  one such cross correlation candidate - that between the 
gravitational 
lensing of the CMB and the flux fluctuations in the  \Lya forest. \par
The gravitational lensing of the CMB, or CMB lensing in short,  is caused by the 
deflection of the CMB photons by the large scale structure potentials  \citep[see][for a
  review]{lewis.challinor:2006}. On large scales, WMAP measurements imply that the primoridal CMB is well described as an isotropic Gaussian random field \cite{Komatsu:2008hk}.  On small scales, 
lensing breaks this isotropy and introduces a specific form of non-Gaussianity. These 
properties of the lensed CMB sky can be used to construct estimators of the deflection 
field that lensed the CMB. Therefore, CMB lensing provides us with a way of 
reconstructing a line-of-sight (los) projected density field from zero redshift to the last 
scattering surface, with a broad geometrical weighting kernel that gets most of its 
contribution from the $z=1-4$ range \cite{hu.okamoto:2002,hirata.seljak:2003,yoo.zaldarriaga:2008}. While CMB lensing is mainly sensitive 
to the geometry and large scale projected density fluctuations, the \Lya forest, the 
absorption in quasar (QSO) spectra caused by intervening neutral hydrogen in the intergalactic 
medium, primarily traces the small-scale distribution of gas (and hence, also matter) 
along the line of sight. 

A cross-correlation between these two effects gives us 
a unique way to study how small scale  fluctuations in the density field evolve on top of 
large scale over and under-densities, and how gas traces the underlying dark matter. 
This signal is therefore a useful tool to test to what extent the fluctuations in the 
\Lya flux relate to the underlying dark matter. Once that relationship is understood, it can also become a powerful probe of the growth of structure on a wide range 
of scales. Since both massive neutrinos and dark energy alter the growth rate of structure at $z\sim 2$, these measurements can probe their effects. This new cross-correlation signal, should also be compared with other existing cross-correlations between CMB and LSS that have already been observed and that are sensitive to different redshift regimes \cite{peiris.spergel:2000, giannantonio08, hirata08, croft06, Xia:2009dr}.

In this work, we build an analytic framework based on simplifying assumptions to estimate the
cross-correlation of the first two moments of the \Lya flux
fluctuation with the weak lensing convergence $\kappa$, obtained from CMB lensing reconstruction, measured along
the same line of sight. The finite resolution of the spectrogram limits the range of parallel $k$-modes probed by the absorption spectra and the finite resolution of the
CMB experiments limits the range of perpendicular
$k$-modes probed by the convergence measurements. These two effect break  the spherical symmetry of the $k$-space integration. However, we show that by resorting to a power series expansion it is still possible to obtain computationally efficient expressions for the evaluation of the signal.

We then investigate the detectability of the 
signal in upcoming CMB and LSS surveys, and the extent to which such a signal can be 
used as a probe of neutrino masses and early dark energy scenarios. A highlight of our results is that the estimated cross-correlation signal 
seems to have significant sensitivity to the normalization of the matter power spectrum $\sigma_8$. Consistency with CMB measurements -- linking power spectrum normalization and the sum of the neutrino masses -- allows to use this cross-correlation to put additional constrain on the latter.
\par

The structure of the paper is as follows. In
Section \ref{sectI} we  introduce the two physical observables,
the \Lya flux and the CMB convergence (\ref{sectIa}), the
cross-correlation estimators (\ref{sectIb}) and their variances (\ref{variance}).  Our
main result is presented in section \ref{sn} where the signal-to-noise
ratios are computed. Section \ref{spectral} contains a spectral
analysis of the observables that aims at finding the \Lya wavenumbers that
contribute most to such a signal. We focus on two cosmologically
relevant applications in sections \ref{neutrinos} and \ref{ede},
for massive neutrinos and early dark energy models, respectively. We
conclude with a discussion in section \ref{discuss}.

\section{Analytical Results}
\label{sectI}
\subsection{Physical Observables}
\label{sectIa}
\subsubsection*{Fluctuations in the \Lya flux}

Using the \textit{fluctuating Gunn--Peterson approximation}
\cite{Gunn:1965hd}, the transmitted flux $\flux$ along a los $\vn$ is
related to the density fluctuations of the intergalactic medium (IGM)
$\delta_{\textrm{IGM}}$ by
\begin{equation}
 \flux(\vn,z)=\exp\left[-A\left(1+\delta_{\textrm{IGM}}(\vn,z)\right)^{\beta}\right],\label
{eq:FGPA}
\end{equation} 
where $A$ and $\beta$ are two functions relating the flux fluctuation
to the dark matter overdensities. These two functions depend on the
redshift considered: $A$ is of order unity and is related to the mean
flux level, baryon fraction, IGM temperature, cosmological parameters
and the photoionization rate of hydrogen. A good approximation for its
redshift dependence is $A(z)\approx 0.0023\,(1+z)^{3.65}$ (see
\cite{kim07}).  $\beta$ on the other hand depends on to the so-called
IGM temperature-density relation and in particular on the power-law
index of this relation (e.g. \cite{huignedin97,mcdonald03}) and should
be less dependent on redshift (unless temperature fluctuations due for
example to reionization play a role, see \cite{mcquinn09}). For the
calculation of signal/noise in the paper, we neglect the
evolution of $A$ and $\beta$ with redshift. While the value of the correlators considered will 
depend on $A$ and $\beta$, their signal-to-noise (S/N) ratio will not.

On scales larger than about $1\,\hMpc$ (comoving), which is about the Jeans
length at $z=3$, the relative \textit{fluctuations} in the \Lya flux
$\delta\flux\equiv(\flux-\bar{\flux})/\bar{\flux}$ are proportional to
the fluctuations in the IGM density field
\cite{Bi:1996fh,croft98,croftweinberg02,Viel:2001hd,saitta08}. We assume that the IGM traces the dark matter on large scales,
\begin{eqnarray}
 \delta\flux(\vn,\chi)&\approx& -A\beta\delta_{\rm IGM}(\vn,\chi)
 \approx -A\beta\delta(\vn,\chi) .\label{deltaF}
\end{eqnarray} 
The (variance of the) flux fluctuation in the redshift
range covered by the \Lya spectrum is then proportional to (the
variance of) the fluctuations in dark matter
\begin{eqnarray}
 \delta\flux^r(\vn)&=&\int_{\chi_i}^{\chi_Q}d\chi\, \delta\flux^r(\vn,\chi)\vs
&\approx&
  \int_{\chi_i}^{\chi_Q}d\chi\, \left(-A\beta\right)^r\delta^r(\vn,\chi),\label{deltaF2}
\end{eqnarray}
where the range of comoving distances probed by the \Lya spectrum
extends from $\chi_i$ to $\chi_Q$. The $r=1$ case corresponds to the
fluctuations in the flux and the $r=2$ case corresponds to their
variance.  We stress that the above approximation is valid in linear
theory neglecting not only the non-linearities produced by
gravitational collapse but also those introduced by the definition of
the flux and those produced by the thermal broadening and peculiar
velocities. Note that while the assumption of ``tracing'' between gas
and dark matter distribution above the Jeans length is 
expected in the standard linear perturbation theory
\cite{eisensteinhu98}, the one between the flux and the matter has
been verified a-posteriori using semi-analytical methods
(\cite{Bi:1996fh,zaroubi06}) and numerical simulations
(\cite{gnedinhui98,croft98,vhs06}) that successfully reproduce most of
the observed \Lya properties. Furthemore, non-gravitational processes
such as temperature and/or ultra-violet fluctuations in
the IGM should alter the \Lya forest flux power and correlations in a
distinct way as compared to the gravitational instability process
and to linear evolution (e.g. \cite{fangwhite04,croft04,slosar09}).

\subsubsection*{Cosmic Microwave Background convergence field}

The effective weak lensing convergence $\kappa(\vn)$ measured along a
los in the direction $\vn$ is proportional to the dark matter
overdensity $\delta$ through
\begin{equation}
 \kappa(\hat{n},\chi_F)=\frac{3H_0^2\Omega_m}{2c^2}\int_0^{\chi_F}d\chi\,W_L
(\chi,\chi_F)\frac{\delta(\hat{n},\chi)}{a(\chi)}\label{kappa},
\end{equation}
where the integral along the los extends up to a comoving distance
$\chi_{F}$ and where $W_L(\chi,\chi_{F})=\chi(\chi_{F}-\chi)/\chi_{F}$
is the lensing window function. In what follows we consider the
cross-correlation of \Lya spectra with the convergence field measured
from the CMB, as in Vallinotto et al.~\cite{Vallinotto:2009wa}, in
which case $\chi_F$ is the comoving distance to the last scattering
surface. Note however that it is straightforward to extend the present
treatment to consider the cross-correlation of the \Lya flux
fluctuations with convergence maps constructed from other data sets,
like optical galaxy surveys.

It is necessary to stress here that Eq.~(\ref{eq:FGPA}) above depends
on the density fluctuations in the IGM, which in principle are
distinct from the ones in the dark matter, whereas $\kappa$ depends on
the dark matter overdensities $\delta$. If the IGM and dark matter
overdensity fields were completely independent, the cross-correlation
between them would inevitably yield zero. If however the fluctuations
in the IGM and in the dark matter are related to one another, then
cross-correlating $\kappa$ and $\delta\flux$ will yield a non-zero
result. The measurement of these cross-correlations tests whether the IGM is tracing the
underlying dark matter field and quantifies the bias between flux and matter.

\subsection{The Correlators}
\label{sectIb}

\subsubsection*{Physical Interpretation}

The two correlators  $\langle
\delta\flux \kappa\rangle$ and $\langle \delta\flux^2 \kappa\rangle$
have substantially different physical meaning:
$\kappa$ is proportional to the over(under)density integrated along
the los and
is dominated by long wavelength modes with $k\sim 10^{-2}$
$\hMpcI$. Intuitively $\kappa$ therefore measures whether a specific
los is probing an overall over(under)dense region. If the IGM traces
the dark matter field, then by Eq.~(\ref{deltaF2}) $\delta\flux$ is
expected to measure the dark matter overdensity along the same los
extending over the redshift range $\Delta z$ spanned by the QSO
spectrum. This implies that
\begin{itemize}
\item $\langle \delta\flux \kappa\rangle$ quantifies
whether and how much the overdensities traced by the \Lya flux
contribute to the overall overdensity measured all the way to the
last scattering surface. Because both $\kappa$ and
$\delta\flux$ are proportional to $\delta$, it is reasonable to expect
that this correlator will be dominated by modes with wavelengths of
the order of hundreds of comoving Mpc. As such, this correlator may be
difficult to measure as it may be more sensitive to the calibration of
the \Lya forest continuum.
\item
$\langle\delta\flux^2\kappa\rangle$ measures the relationship between long wavelength modes in the density and the amplitude of the variance of the flux. The variance on small scales and the amplitude of fluctuations on large-scales are not coupled in linear theory. However, in non-linear gravitational theory regions of higher mean density have higher matter fluctuations.  These lead to higher amplitude fluctuations in flux \cite{2001ApJ...551...48Z}. Since
$\langle\delta\flux^2\kappa\rangle$ is sensitive to this interplay
between long and short wavelength modes, this correlator is much more
sensitive than $\langle \delta\flux \kappa\rangle$ to the structure
growth rate. Furthermore, because $\delta\flux^2$ is sensitive to short
wavelengths, this signal is dominated
by modes with shorter wavelength than the ones dominating
$\langle \delta\flux \kappa\rangle$. As such, this signal should
be less sensitive to the fitting of the continuum of the \Lya forest.
\end{itemize}

\subsubsection*{Tree level approximation}

In what follows we focus on obtaining analytic expressions for the correlations 
between the (variance of the) flux
fluctuations in the \Lya spectrum and the CMB convergence $\kappa$
measured along the same los. From  Eqs.~(\ref{deltaF2}, \ref{kappa})
above it is straightforward to obtain the general expression for the signal
\begin{align}\label{eq:deltaFmK_1}
 \langle \delta\flux^r(\hat{n}) \kappa(\hat{n})\rangle&=\frac{3H_0^2\Omega_m}
{2c^2}\int_0^{\chi_F}d\chi_c \frac{W_L(\chi_c,\chi_F)}{a(\chi_c)} \vs
&\times\int_{\chi_i}^{\chi_Q}d\chi_q \,(-A\beta)^r\,
\langle \delta^r(\hat{n},\chi_q) \, \delta(\hat{n},\chi_c) \rangle.
\end{align}
Since the QSOs used to measure the \Lya forest lie at $z>2$, it is
reasonable to expect that non-linearities induced by gravitational
collapse will not have a large impact on the final results. In the following we
therefore calculate the $r=1$ and $r=2$ correlators at
\textit{tree-level} in cosmological perturbation theory. While beyond the scope of the current calculation, we could include the effects of non-linearities induced by gravitational collapse by applying the \textit{Hyperextended
  Perturbation Theory} of Ref.~\cite{Scoccimarro:2000ee} to the terms in Eq.~(\ref{eq:deltaFmK_1}).

At tree level in perturbation theory the redshift dependence of the
matter power spectrum factorizes into
$P(k,\chi_c,\chi_q)=P_L(k)\,D(\chi_c)\,D(\chi_q)$, where $P_L(k)$
denotes the zero-redshift linear power spectrum and $D(\chi)$ the
growth factor at comoving distance $\chi$. Furthermore, the correlator
appearing in the integrand of Eq.~(\ref{eq:deltaFmK_1}) depends on the
separation $\Delta\chi=\chi_q-\chi_c$ between the two points running
on the los and in general it will be significantly non-zero only when
$|\Delta\chi|\leq \Delta\chi_0\approx 150 \hMpc$. Also, at tree level
in perturbation theory these correlators carry $2r$ factors of
$D$.\footnote{Notice in fact that even though in the $r=2$ case it
  would be reasonable to expect three factors of $D$, the first
  non-zero contribution to the three-point function carries four
  factors of $D$ because the gaussian term vanishes exactly.} Using
the approximation
\begin{align}
 D(\chi_c)&=&D(\chi_q-\Delta\chi)&\approx D(\chi_q),\label{eq:approx_D}\\
W_L(\chi_c,\chi_F)&=&W_L(\chi_q-\Delta\chi,\chi_F)&\approx W_L(\chi_q,\chi_F),\\
a(\chi_c)&=&a(\chi_q-\Delta\chi)&\approx a(\chi_q),\label{eq:approx_a}
\end{align}
we can then write $\langle \delta^r(\hat{n},\chi_q) \, \delta(\hat{n},\chi_c) \rangle 
\approx \xi_r(\Delta\chi)D^{2r}(\chi_q)$ and trade the double integration (over $
\chi_c$ and $\chi_q$) for the product of two single integrations over $\Delta\chi$ 
and $\chi_q$. Equation (\ref{eq:deltaFmK_1}) factorizes into
\begin{align}\label{eq:deltaFmK_2}
 \langle \delta\flux^r \kappa\rangle&\approx(-A
\beta)^r\frac{3H_0^2\Omega_m}{2c^2}\int_{\chi_i}^{\chi_Q}d\chi_q \frac{W_L(\chi_q,\chi_F)}{a(\chi_q)} D^{2r}
(\chi_q) \vs
&\times\int_{-\Delta\chi_0}^{\Delta\chi_0}d\Delta\chi \,
\xi_r(\Delta\chi).
\end{align}
This is the expression used to evaluate the signal. The determination of an 
expression for $\xi_r$ and of an efficient way for evaluating it is the focus of the 
rest of the section.

\subsubsection*{Window Functions}

The experiments that measure the convergence and the flux fluctuations have finite resolutions. We approximate the effective window functions of 
these experiments by analytically tractable Gaussian function.

These two window functions act differently: the finite resolution of the CMB convergence measurements limits the accessible range of modes perpendicular to the los, $\vec{k}_{\perp}$, and the
finite resolution of the \Lya spectrum limits the range of accessible modes $k_{\parallel}$ parallel to the los.  This
separation of the modes into the ones parallel and perpendicular to
the los is intrinsically dictated by the nature of the observables and
it cannot be avoided once the finite resolution of the various
observational campaigns is taken into account. Because of this symmetry, the calculation is most transparent in cylindrical coordinates: $\vec{k}=k_{\parallel}\vn+\vec{k}_{\perp}$.

The high-$k$ (short wavelength) cutoff scales for the CMB and \Lya
modes are denoted by $k_C$ and $k_L$ respectively. Furthermore, we
also add a low-$k$ (long wavelength) cutoff for the \Lya forest, to
take into account the fact that wavelengths longer than the spectrum
will appear in the spectrum itself as a background. We denote this
low-$k$ cutoff by $k_l$. After defining the auxiliary quantities
\begin{eqnarray}
\bar{k}^2&\equiv&\frac{k_L^2\,k_l^2}{k_L^2+k_l^2},\\
\hat{k}^2&\equiv&\frac{k_L^2\,k_l^2}{2k_l^2+k_L^2},
\end{eqnarray}
the window functions acting on the \Lya and on the CMB modes, denoted
respectively by $W_{\alpha}$ and $W_{\kappa}$, are defined through
\begin{eqnarray}
 W_{\alpha}(k_{\parallel},k_L,k_l)&\equiv&\left[1-e^{-(k_{\parallel}/k_l)^2}\right]e^{-
(k_{\parallel}/k_L)^2}\nonumber\\
 &=&e^{-(k_{\parallel}/k_L)^2}-e^{-(k_{\parallel}/\bar{k})^2},\label{eq:klkL}\\
 W_{\kappa}(\vec{k}_{\perp},k_C)&\equiv&e^{(-\vec{k}^2_{\perp}/k^2_C)},
\end{eqnarray}
where the direction dependence of the two window functions  has been made explicit.

We determine the values of the cutoff scales as follows. For the \Lya
forest, we consider the limitations imposed by the spectrograph,
adopting the two cutoff scales $k_L$ and $k_l$ according to the
observational specifications.  For the reconstruction of the CMB
convergence map we compute the minimum variance
lensing reconstruction noise following Hu and Okamoto
\cite{Hu:2001kj}. We then identify the multipole $l_c$, where the
signal power spectrum equals the noise power spectrum for the
reconstructed deflection field (for $l>l_c$ the noise is higher than
the signal). Finally, we translate the angular cutoff $l_c$ into a 3-D
Fourier mode $k_C$ at the relevant redshift so to keep only modes with
$k\le k_C$ in the calculation. Note that if we had used the shape of
the noise curve instead of this Gaussian cutoff, we would have effectively
retained more Fourier modes, thereby increasing the signal. However,
to keep the calculations simple and conservative we use the above Gaussian window. In what follows, we will present results for convergence
map reconstructions from the datasets of two CMB experiments: Planck
and an hypothetical CMB polarization experiment based on a proposed new camera for the Atacama Cosmology Telescope (ACTPOL). For the
former, we adopt the sensitivity values of the 9 frequency channels
from the Blue Book \cite{PlanckBB}. For the latter we assume a
hypothetical polarization based CMB experiment with a $3$ arcmin beam
and $800$ detectors, each having a noise-equivalent-temperature (NET)
of 300 $\mu$K-$\sqrt{s}$ over $8000$ sq. deg., with an integration
time of $3\times 10^7$ seconds. We further assume that both
experiments will completely cover the $8000$ sq. deg footprint of
BOSS.
\begin{figure*}
\includegraphics[width=0.49\textwidth]{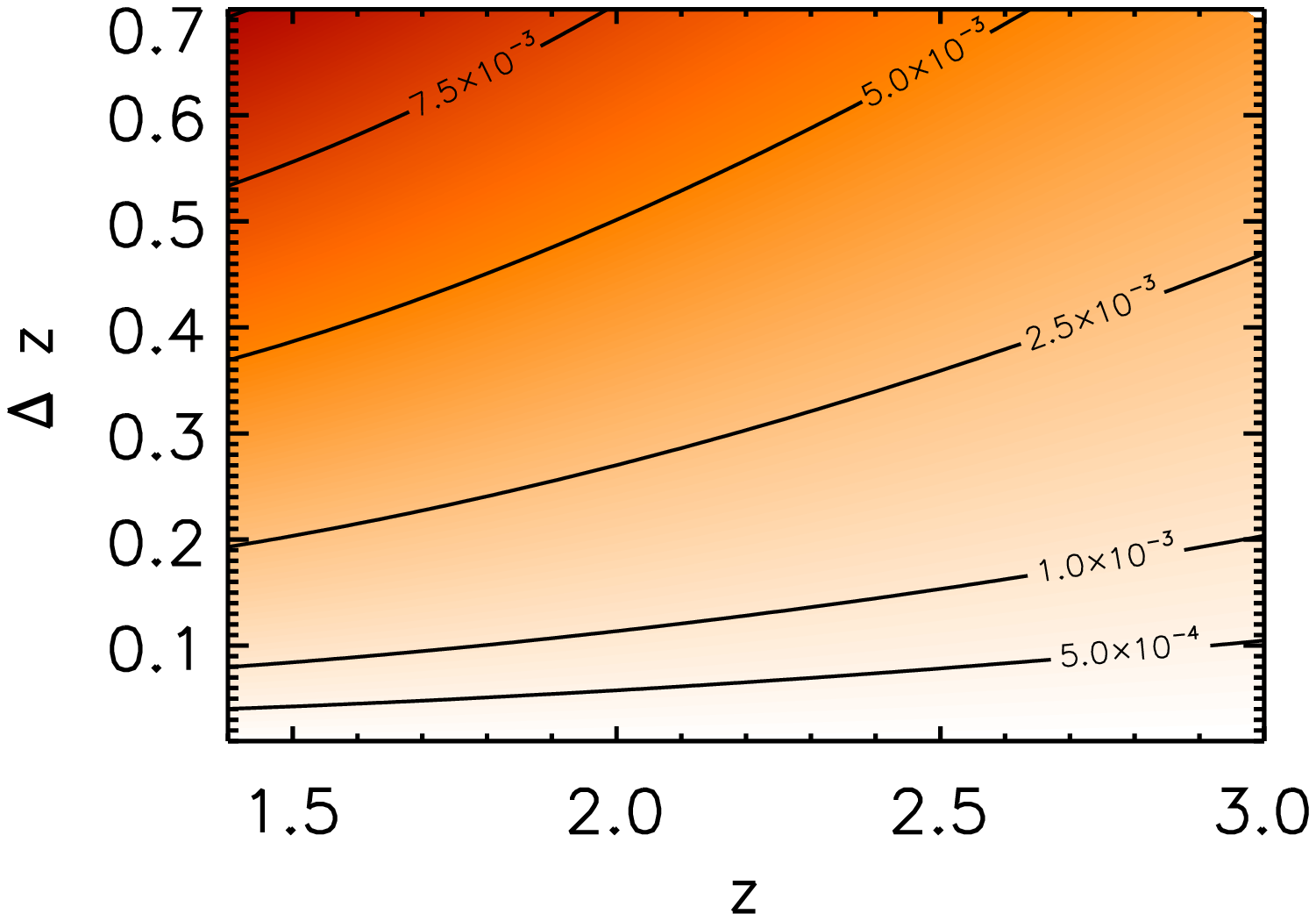}
\includegraphics[width=0.49\textwidth]{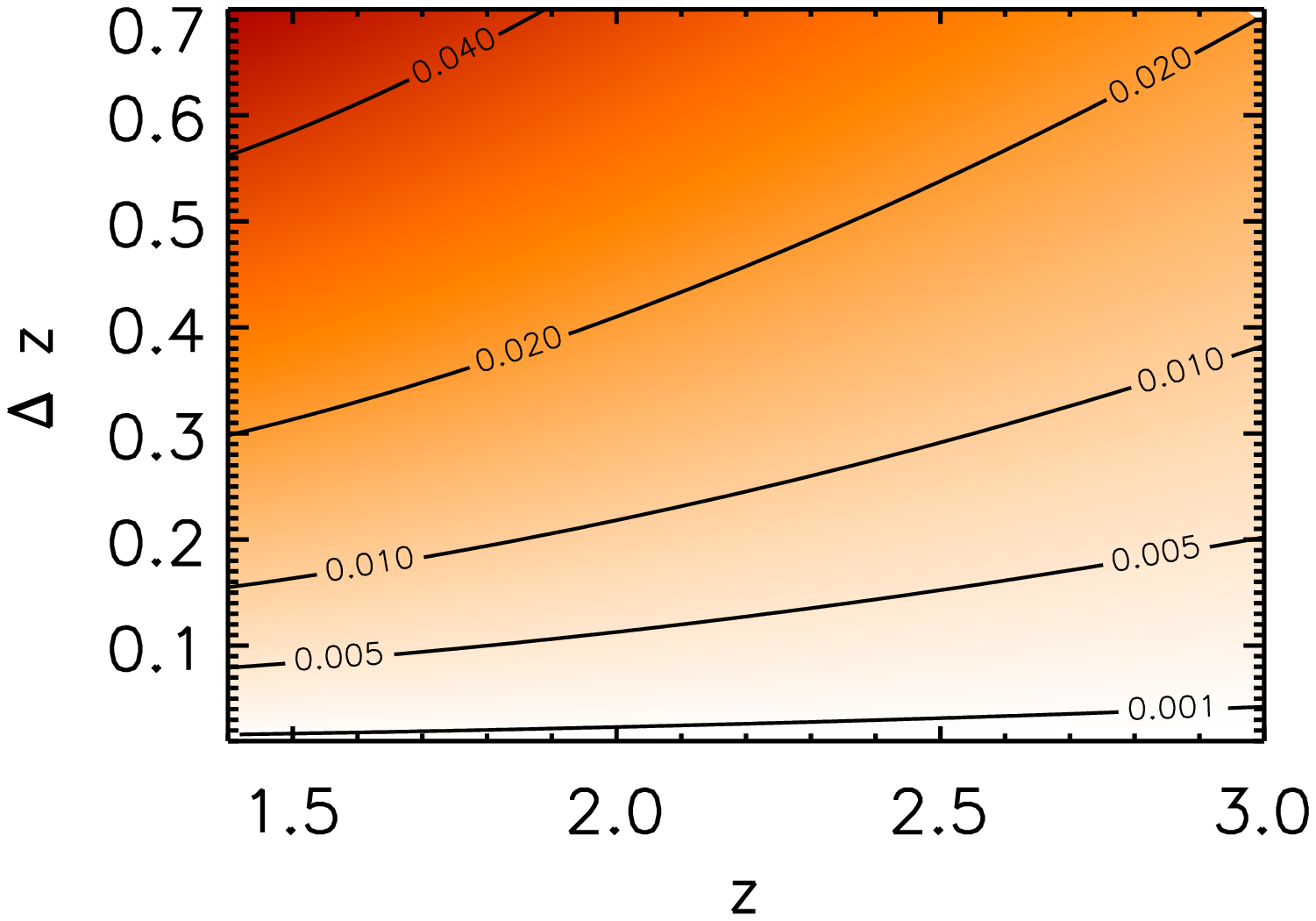}
\caption{\label{Fig:dK_S_3_2} Absolute value of the correlator $\langle
  \delta\flux \kappa\rangle$ along a \textit{single} line-of-sight as
  a function of the source redshift $z$ and of the length of the
  measured spectrum $\Delta z$, for convergence maps recontructed from
  Planck 
(left panel, $k_C=0.021\,\hMpcI$) and ACTPOL (right panel, $k_C=0.064\,\hMpcI$). The value of the
  resolution of the QSO spectrum is the one predicted for SDSS-III,
  $k_L=4.8 \hMpcI$. To make the physics of structure formation
  apparent, we turn off the IGM physics by setting $A=\beta=1$ (it is
  straightforward to rescale the values of the correlator to reflect
  different values of $A$ and $\beta$).}
\end{figure*}

\subsubsection*{Auxiliary Functions}  

Because the calculation has cylindrical rather
than spherical symmetry, the evaluation of the correlators of Eq.~(\ref{eq:deltaFmK_2}) is more complicated, particularly for $r > 1$. As shown in the appendix, it
is possible to step around this complication and to obtain results
that are computationally efficient with the adoption of a few
auxiliary functions that allow the integrations in $k$-space to be
carried out in two steps, first integrating on the modes perpendicular
to the los, and subsequently on the ones parallel to the los.
The
perturbative results for the correlators are expressed as combinations
of the following auxiliary functions:
\begin{widetext}
\begin{align}
 \tilde{H}_m(k_{\parallel};k_C)&\equiv \int_{|k_{\parallel}|}^{\infty}\frac{k\,dk}{2\pi}\,
\frac{P_L(k)}{m!}\,\left(\frac{k^2-k_{\parallel}^2}{k_C^2}\right)^{m}\,\exp\left(-\frac
{k^2-k_{\parallel}^2}{k_C^2}\right), & \label{eq:DefA}\\
\tilde{L}_m(k_{\parallel};k_C)&\equiv
\int_{|k_{\parallel}|}^{\infty}\frac{dk}{2\pi k} \,\frac{P_L(k)}{m!}\,\left(\frac{k^2-k_
{\parallel}^2}{k_C^2}\right)^{m}\,\exp\left(-\frac{k^2-k_{\parallel}^2}{k_C^2}\right),& 
\label{eq:Defbeta}\\
 f^{(n)}_m(\Delta\chi;k_C,k_L)&\equiv \int_{-\infty}^{\infty}\frac{dk_{\parallel}}{2\pi}\,
\left(\frac{k_{\parallel}}{k_L}\right)^n\exp\left[-\frac{k_{\parallel}^2}{k_L^2}+ik_
{\parallel}\Delta\chi\right]\tilde{f}_m(k_{\parallel};k_C) &\textrm{ with } f=\{L,H\},
\label{eq:Deff_m^n}\\
 \bar{f}_0^{(n)}(s)&\equiv \int_{-\infty}^{\infty}\frac{dk}{2\pi}\left(\frac{k}{s}\right)^n
\left[e^{-2k^2/s^2}-e^{-k^2/\hat{k}^2}\right]\tilde{f}_0(k;\infty)  &\textrm{ with } f=\{L,H
\}.\label{eq:Deff_bar}
\end{align}
\end{widetext}
Equations (\ref{eq:DefA}) and (\ref{eq:Defbeta}) above represent an
intermediate step, where the integration on the modes perpendicular to
the los is carried out. Equations (\ref{eq:Deff_m^n}) and
(\ref{eq:Deff_bar}) are then used to carry out the remaining
integration over the modes that are parallel to the los.

The symmetry properties of the auxiliary functions are as follows. The
functions $\tilde{f}_m$ are \textit{real and even} in $k_{\parallel}$
regardless of the actual value of $m$. This in turn implies that
$f^{(n)}_m$ are real and even (imaginary and odd) in $\Delta\chi$ when
$n$ is even (odd). Furthermore, the coefficients $\bar{f}_0^{(n)}$ are real and 
non-zero only if $n$ is even, thus ensuring that $\xi_r(\Delta\chi)$ is always real-valued.

\begin{figure*}
\includegraphics[width=0.49\textwidth]{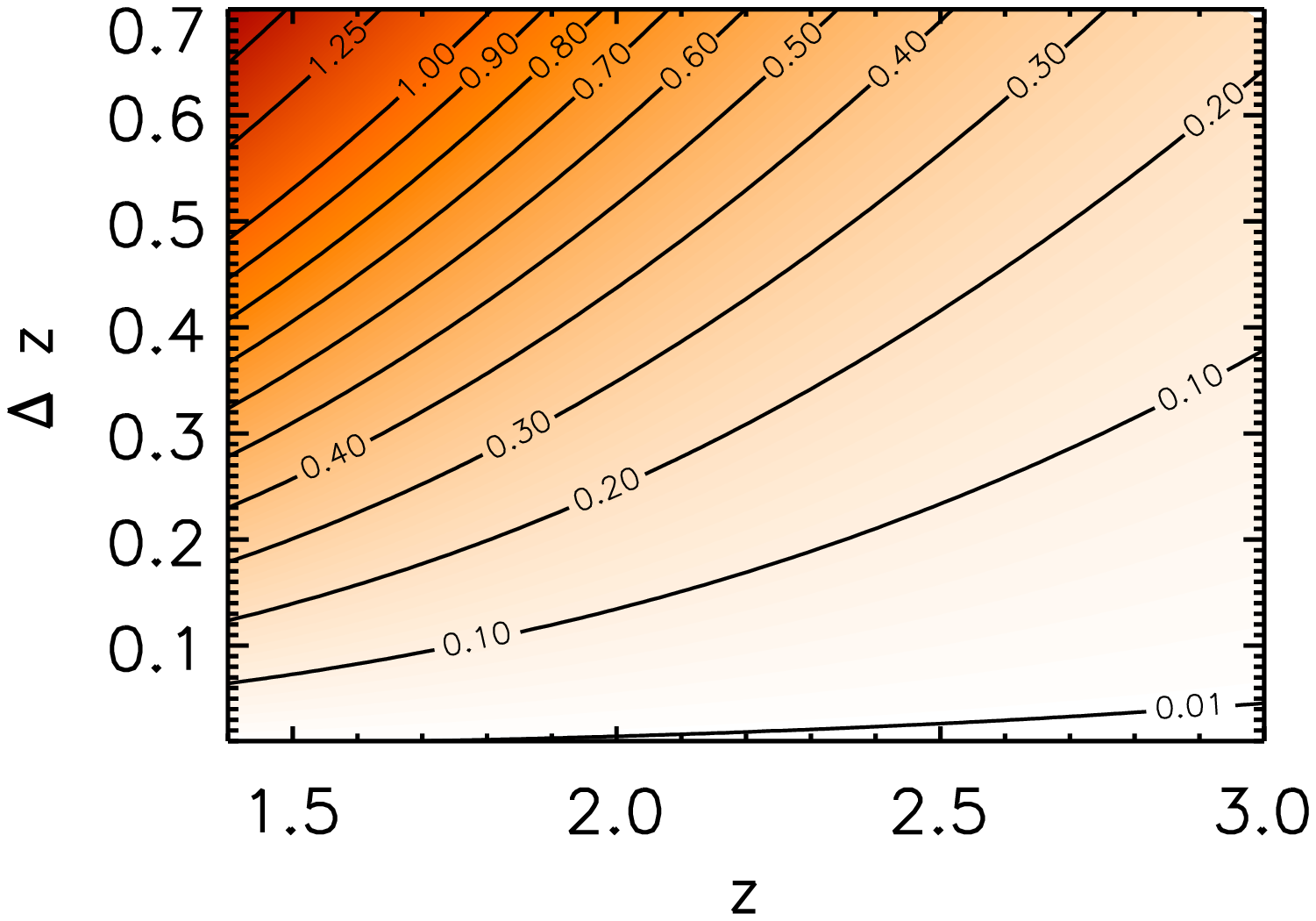}
\includegraphics[width=0.49\textwidth]{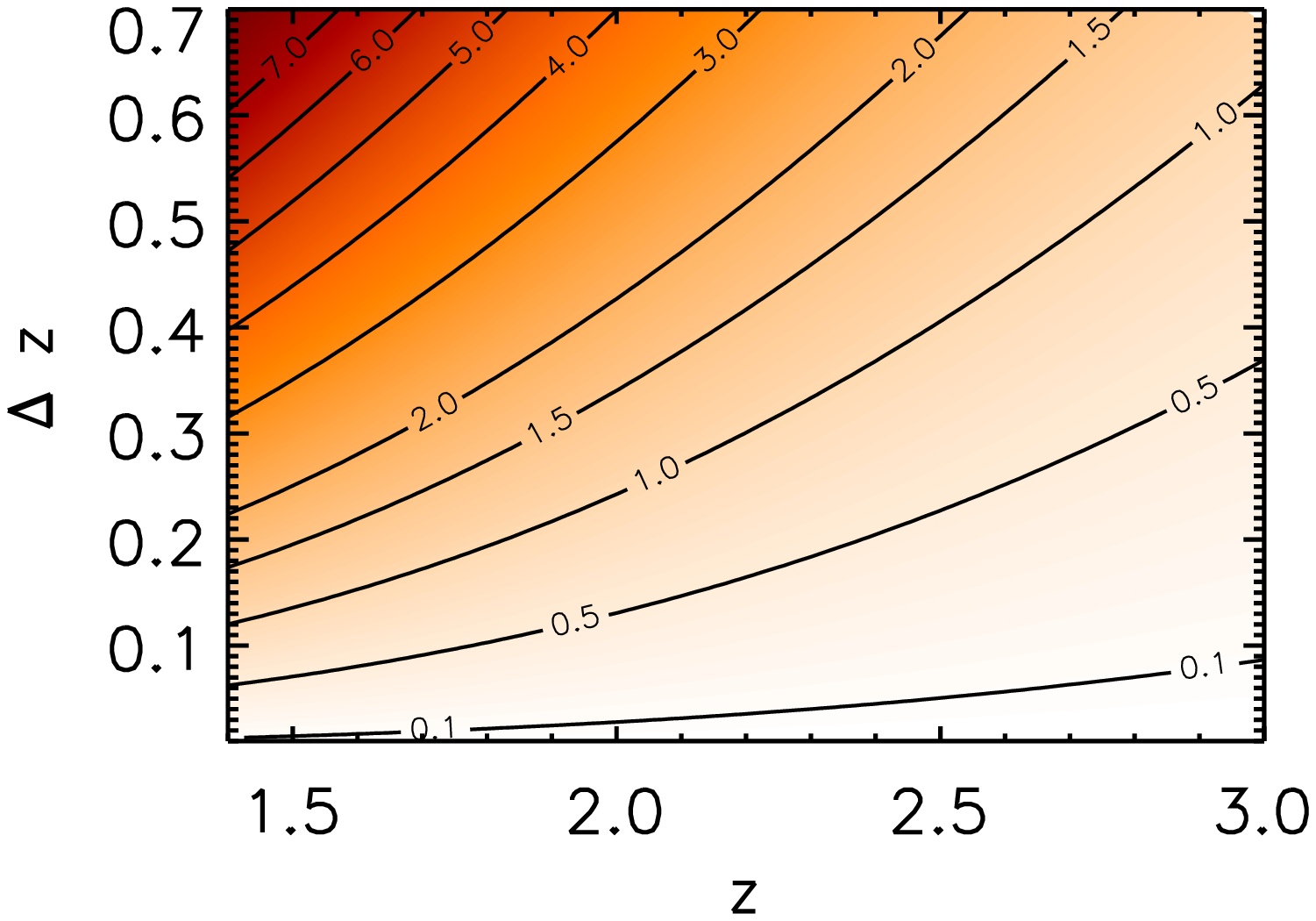}
\caption{\label{Fig:S_3_2} Absolute value of the correlator $\langle
  \delta\flux^2 \kappa\rangle$ along a \textit{single} line-of-sight
  as a function of the source redshift $z$ and of the length of the
  measured spectrum $\Delta z$, for convergence maps recontructed from Planck 
(left panel, $k_C=0.021\,\hMpcI$) and ACTPOL (right panel, $k_C=0.064\,\hMpcI$). The value of the resolution of the QSO 
spectrum is the one predicted for SDSS-III, $k_L=4.8 \hMpcI$. As before, we set 
$A=\beta=1$ to make the physics of structure formation apparent.}
\end{figure*}

\subsubsection*{The $\langle \delta\flux \kappa\rangle$ correlator}

In the $r=1$ case it is straightforward to identify
$\xi_1(\Delta\chi)$ with a two point correlation function measured
along the los. However, the intrinsic geometry of the problem and the
inclusion of the window functions leads to evaluate this correlation
function in a way that is different from the usual case, where the
spherical symmetry in $k$-space can be exploited. In the present case
we have
\begin{equation}
 \xi_1(\Delta\chi)=H_0^{(0)}(\Delta\chi;k_C,k_L)- H_0^{(0)}(\Delta\chi;k_C,\bar{k}).
\label{eq:xi_1}
\end{equation} 
It is then straightforward to plug
Eq.~(\ref{eq:xi_1}) into Eq.~(\ref{eq:deltaFmK_2}) to obtain $\langle
\delta\flux(\hat{n}) \kappa(\hat{n})\rangle$.\footnote{We checked that in the limit where $k_L\rightarrow\infty$, $k_C
\rightarrow\infty$ and $k_l\rightarrow 0$ the usual two point correlation function is 
recovered. Whereas one would naively expect that letting $k_L=k_C$ and 
$k_l=0$ would lead to recover the usual two point function calculated exploiting 
spherical symmetry in $k$-space with a cutoff scale equal to the common $k_L$, 
this is actually \textit{not} the case. The reason for this is that the volume of $k$-
space over which the integration is carried out is different for the two choices of 
coordinate systems. In particular, the spherical case always includes fewer modes 
than the cylindrical one. The two results therefore coincide \textit{only} in the $k_L
\rightarrow \infty$ limit.}
In Fig.~\ref{Fig:dK_S_3_2} we show the absolute value of the cross-correlation of the
convergence $\kappa$ of the CMB with the \Lya flux fluctuations
$\delta\flux$ observed for a quasar located at redshift $z$ and whose
spectrum spans a range of redshift $\Delta z$. The cosmological model
used (and assumed throughout this work) is a flat \lcdm universe with
$\Omega_{\rm m}=0.25$, $h=0.72$ and $\sigma_8=0.84$ consistent with
the WMAP-5 cosmology \cite{Komatsu:2008hk}. The left and right panel
show the results for the resolution of Planck and of the proposed ACTPOL 
experiment. We artificially set
$A=\beta=1$, effectively ``turning off'' the physics of IGM: this
choice is not dictated by any physical argument but from the fact that
it makes apparent the dynamics of structure formation. 

The behavior of $\langle \delta\flux\, \kappa\rangle$ shown in
Fig.~\ref{Fig:dK_S_3_2} makes physical sense. Recall that this
correlator is sensitive to the overdensity integrated along the
redshift interval $\Delta z$ (spanned by the QSO spectrum) that
contributes to the CMB convergence. It then increases almost linearly
with the length of the QSO spectrum $\Delta z$. It also increases
if the resolution of the CMB experiment $k_C$ is increased.  An
increased value of $\Delta z$ corresponds to a longer \Lya spectrum,
carrying a larger amount of information and thus leading to a larger
correlation. Similarly, an increased value of $k_C$ corresponds to a
higher resolution of the reconstructed convergence map and therefore
more modes -- and information -- being included in the
correlation. Deepening the source's redshift (while keeping $A$ and
$\beta$ fixed) on the other hand results in a \textit{decrease} in
$\langle \delta\flux\, \kappa\rangle$. This fact is related to the
growth of structure: the spectrum of a higher redshift QSO is probing
regions where structure is less clumpy and therefore the absolute
value of the correlation is smaller. Finally, once the redshift
dependence of $A$ is turned on ($\beta$ is only mildly redshift
dependent) the above result change, leading to a final signal that is
increasing with redshift.

We stress here
that values of the correlators will be
different when $A$ and $\beta$ are different from unity. Ultimately these values 
should be recovered from a full non-linear
study based on large scale-high resolution hydrodynamical
simulations. However, numerical studies based on hydrodynamical
simulations have shown convincingly that for both the flux power
spectrum (2-pt function) and flux bispectrum (3-pt function) the shape
is very similar to the matter power and bispectrum, while the
amplitude is usually matched for values of $A$ and $\beta$ that are
different from linear predictions (see discussion in
\cite{vielbisp}). In this framework, non-linear hydrodynamical
simulations should at the end provide the ``effective'' values for $A$
and $\beta$ that will match the observed correlators and our results
can be recasted in terms of these new parameters in a straightforward
way.

\subsubsection*{The $\langle \delta\flux^2 \kappa\rangle$ correlator}

The $r=2$ case, where the variance of the flux fluctuation
$\delta\flux^2$ integrated along the los is cross-correlated with $\kappa$, is more
involved. Looking back at Eqs.~(\ref{eq:deltaFmK_1},
\ref{eq:deltaFmK_2}) it is possible to realize that the cumulant
correlator $\langle \delta^2(\hat{n},\chi_q) \, \delta(\hat{n},\chi_c)
\rangle=\xi_2(\Delta\chi)$ corresponds to a collapsed three-point
correlation function, as two of the $\delta$'s refer to the same
physical point. The evaluation of $\xi_2$ is complicated by the
introduction of the window functions $W_{\alpha}$ and
$W_{\kappa}$. For sake of clarity, we report here only the final
results at tree level in cosmological perturbation theory, relegating
the lengthy derivation to the appendix. Letting
\begin{equation}
 \xi_2(\Delta\chi)=\langle\delta^2_q\delta_c\rangle_{1,2}
+2\langle\delta^2_q\delta_c\rangle_{2,3},\label{eq:Def_ddd12_ddd23}
\end{equation}  
and using the auxiliary functions defined in Eqs.~(\ref{eq:DefA}-\ref{eq:Deff_bar}) 
above, it is possible to obtain the following series solution
\begin{widetext}
\begin{eqnarray}
   \langle\delta^2\delta\rangle_{1,2}&=&
2\sum_{m=0}^{\infty}\left\{\frac{5}{7}\,
\left[H_m^{(0)}(\Delta\chi,\chi_q;k_C,k_L)-H_m^{(0)}(\Delta\chi,\chi_q;k_C,\bar{k})
\right]^2\right.\nonumber\\
&+&\left[k_L\,H_m^{(1)}(\Delta\chi,\chi_q;k_C,k_L)-\bar{k}\,H_m^{(1)}(\Delta\chi,
\chi_q;k_C,\bar{k})\right]
\left[k_L\,L_m^{(1)}(\Delta\chi, \chi_q;k_C,k_L)-\bar{k}L_m^{(1)}(\Delta\chi, 
\chi_q;k_C,\bar{k})\right]
\vs
&-&m\,k_C^2\,\left[H_m^{(0)}(\Delta\chi,\chi_q;k_C,k_L)-H_m^{(0)}(\Delta\chi,
\chi_q;k_C,\bar{k})\right]\left[L_m^{(0)}(\Delta\chi,\chi_q;k_C,k_L)-L_m^{(0)}(\Delta
\chi,\chi_q;k_C,\bar{k})\right]
\vs
&+&
\frac{2}{7}\left[k_L^2\,L_m^{(2)}(\Delta\chi,\chi_q;k_C,k_L)-\bar{k}^2\,L_m^{(2)}
(\Delta\chi,\chi_q;k_C,\bar{k})\right]^2\vs
&-&\frac{4m}{7}\,
k_C^2\,\left[k_L\,L_m^{(1)}(\Delta\chi,\chi_q;k_C,k_L)-\bar{k}\,L_m^{(1)}(\Delta\chi,
\chi_q;k_C,\bar{k})\right]^2
\vs
&+&\left.\frac{m(2m-1)}{
7}\,k_C^4\,\left[L_m^{(0)}(\Delta\chi,\chi_q;k_C,k_L)-L_m^{(0)}(\Delta\chi,
\chi_q;k_C,\bar{k})\right]^2
\right\},\label{d2d_12_kLkl_final}\\
 \langle\delta_q^2\delta_c\rangle_{2,3}&=&
2\sum_{m=0}^{\infty}\frac{(-1)^m\,2^m}{m!}
\left[\frac{6}{7}\bar{H}_0^{(m)}(k_L)H_0^{(m)}(\Delta\chi;k_C,k_L)\right.
+\frac{1}{2}k_L^2\bar{L}_0^{(m+1)}(k_L)\,H_0^{(m+1)}(\Delta\chi;k_C,k_L)\vs
&+&\frac{1}{2}k_L^2\bar{H}_0^{(m+1)}(k_L)\,L_0^{(m+1)}(\Delta\chi;k_C,k_L)
+\frac{3}{7}k_L^4\,\bar{L}_0^{(m+2)}(k_L)\,L_0^{(m+2)}(\Delta\chi;k_C,k_L)\vs
&-&\left.
\frac{k_L^2}{7}
\bar{H}_0^{(m)}(k_L)\,L_0^{(m+2)}(\Delta\chi;k_C,k_L)
-\frac{k_L^2}{7}
\bar{L}_0^{(m+2)}(k_L)\,H_0^{(m)}(\Delta\chi;k_C,k_L)
+ (k_L\rightarrow\bar{k})\right].\label{eq:d2d_23final_kLkl}
 \end{eqnarray}
\end{widetext}

In Fig.~\ref{Fig:S_3_2} we show the result obtained using the tree
level expression for $\langle\delta\flux^2\kappa\rangle$,
Eqs.~(\ref{eq:Def_ddd12_ddd23}-\ref{eq:d2d_23final_kLkl}). As before,
we focus on the physics of structure formation and we turn off the IGM
physics by setting $A=\beta=1$. First, it is necessary to keep in mind
that $\langle\delta\flux^2\kappa\rangle$ is sensitive to the interplay
of long and short wavelength modes and it probes the enhanced growth of
short wavelength overdensities that lie in an environment
characterized by long wavelength overdensities. The behavior of
$\langle\delta\flux^2\kappa\rangle$ with respect to $z$ and $\Delta z$
is similar to that of $\langle\delta\flux\kappa\rangle$: it increases
if $\Delta z$ is increased or if the QSO redshift is
decreased. However, the effect of the growth of structure is in this
case stronger than in the previous case. This does not come as a
surprise, as the growth of structure acts coherently in two ways on
$\langle\delta\flux^2\kappa\rangle$. Since in a \lcdm model all modes
grow at the same rate, a lower redshift for the source QSO implies
larger overdensities on large scales which in turn enhance even
further the growth of overdensities on small scales. Thus by lowering
the source's redshift two factor play together to enhance the signal:
first the fact that long and short wavelength modes have both grown
independently, and second the fact that being coupled larger
long-wavelength modes boost the growth of short wavelength modes by a
larger amount. This dependence is also made explicit in
Eq.~(\ref{eq:deltaFmK_2}), where we note that
$\langle\delta\flux^2\kappa\rangle$ depends on four powers of the
growth factor. Finally, as before, the higher the resolution of the
CMB experiment the larger is
$d\langle\delta\flux^2\kappa\rangle/d\Delta z$. This too makes
physical sense, as a larger resolution leads to more modes
contributing to the signal and therefore to a larger
cross-correlation.

\subsection{Variance of correlators}
\label{variance}
To assess whether the correlations between fluctuations in the flux and
convergence are detectable we need to estimate the signal-to-noise
ratio, which in turn requires the evaluation of the noise associated
with the above observable. As mentioned above, both instrumental noise
and cosmic variance are considered. We then move to estimate the
variance of our correlator
\begin{eqnarray}
 \sigma_r^2\equiv\langle\delta\flux^{2r}\kappa^2\rangle-\langle\delta\flux^{r}
\kappa\rangle^2.
\end{eqnarray} 
Since $\langle\delta\flux^{r}\kappa\rangle^2$ is just the square of
the signal, we aim here to obtain \textit{estimates} for
$\langle\delta\flux^{2r}\kappa^2\rangle$. From
Eq.~(\ref{eq:deltaFmK_1}), we get:
\begin{align}
 \langle \delta\flux^{2r} \kappa^2\rangle&=\left(A^r\beta^r \frac{3H_0^2\Omega_m}
{2c^2}\right)^2\int_0^{\chi_F}d\chi_c \frac{W_L(\chi_c,\chi_F)}{a(\chi_c)}\vs
&\times
\int_0^{\chi_F}d\chi'_c \frac{W_L(\chi'_c,\chi_F)}{a(\chi'_c)} 
\int_{\chi_i}^{\chi_Q}d\chi_q \,\int_{\chi_i}^{\chi_Q}d\chi'_q
\vs
&\times
\langle \delta^{r}(\hat{n},\chi_q) \,\delta^{r}(\hat{n},\chi'_q)\, \delta(\hat{n},\chi_c)
\delta(\hat{n},\chi'_c) \rangle \, ,
\label{eq:deltaF2mK2_1}
\end{align}
where there are now two integrals running along the convergence los (on
$\chi_c$ and $\chi_c'$) and two running along the \Lya spectrum (on
$\chi_q$ and $\chi_q'$).  The correlator appearing in the integrand of
Eq.~(\ref{eq:deltaF2mK2_1}) is characterized by an even ($2r+2$)
number of $\delta$ factors. This implies that an approximation to its
value can be obtained using Wick's theorem. When
Wick's theorem is applied, many different terms will in general
appear. Adopting for sake of brevity the 
notation $\delta(\hat{n},\chi'_i)\equiv\delta_i$, terms characterized by the contraction of
$\delta_i$ and $\delta_j$ will receive non-negligible contributions
over the overlap of the respective los. The terms providing the
largest contribution to $\langle \delta\flux^{2r} \kappa^2\rangle$ are
the ones where $\delta_c$ is contracted with
$\delta_{c'}$: these terms in fact contain the value of the
cosmic variance of the convergence and receive significant contributions
from all points along the los from the observer all the way to the
last scattering surface. On the other hand, whenever we consider the
cross-correlation between a $\delta_c$ and a $\delta_q$, this will
acquire a non-negligible value only for those set of points where
the los to the last scattering surface overlaps with the \Lya
spectrum. As such, these terms are only proportional to the length of
the \Lya spectrum, and thus sensibly smaller than the ones containing
the variance of the convergence. We note in passing that the same argument should also apply to the connected part of the correlator, which should be significantly non-zero only along the \Lya spectrum. Mathematically, these facts become
apparent from Eq.~(\ref{eq:deltaF2mK2_1}) above, where terms
containing $\langle\delta_c\delta_{c'}\rangle$ are the only ones for
which the integration over $\chi_c$ and $\chi'_{c}$ can be traded for
an integration over $\Delta\chi_c$ \textit{and} an integration over
$\chi_c$ that extends \textit{all the way to} $\chi_F$. If on the
other hand $\delta_c$ is contracted with a $\delta_q$ factor, then the
approximation scheme of Eqs.~(\ref{eq:approx_D}-\ref{eq:approx_a})
leads to an integral over $\Delta\chi$ and to an integral over
$\chi_q$ that extends only over the length probed by the \Lya
spectrum. It seems therefore possible to safely neglect terms where the $\delta$'s
referring to the convergence are not contracted with each other.

\subsubsection*{The variance of $\delta\flux\,\kappa$}

\begin{figure*}
\includegraphics[width=0.49\textwidth]{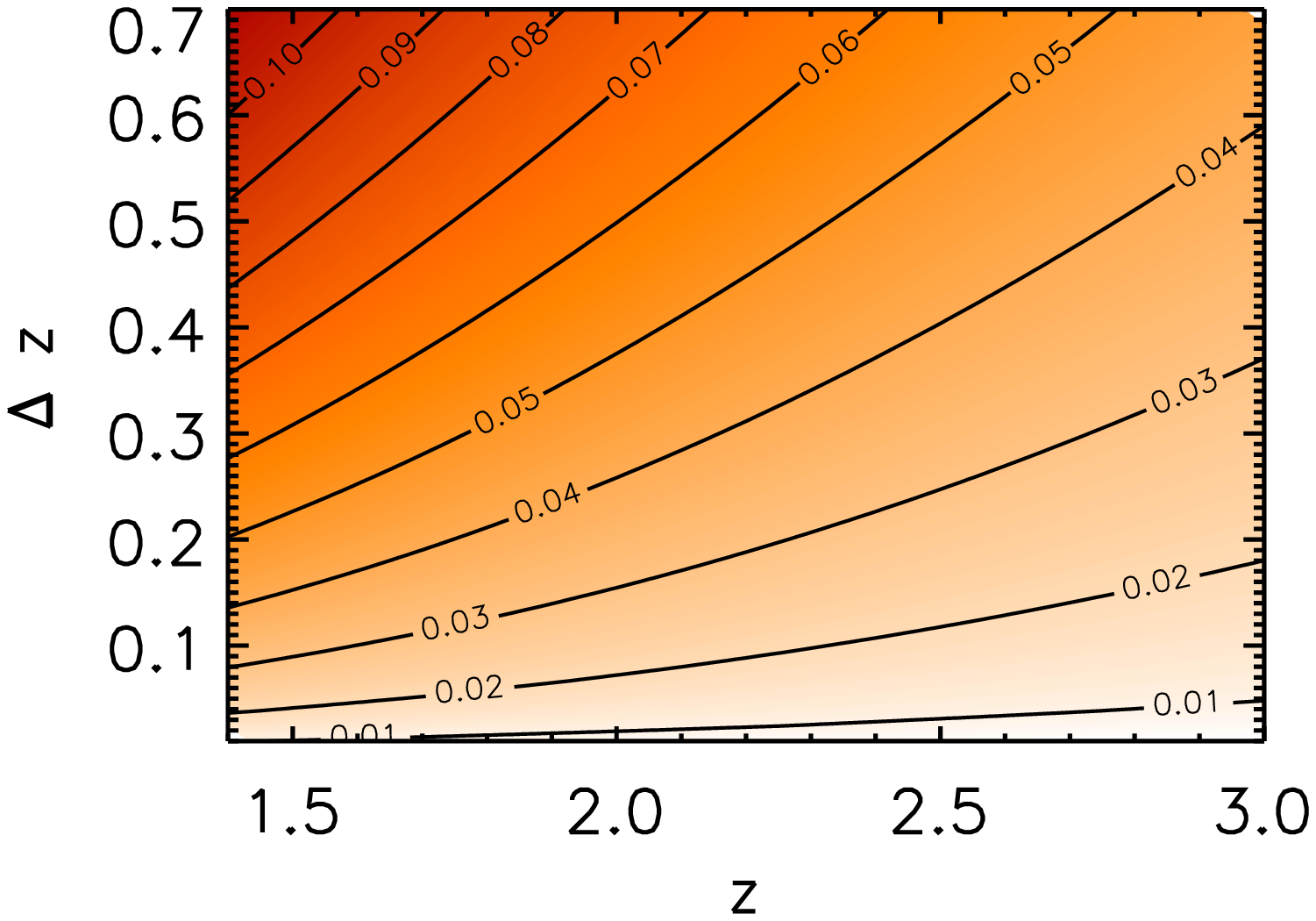}
\includegraphics[width=0.49\textwidth]{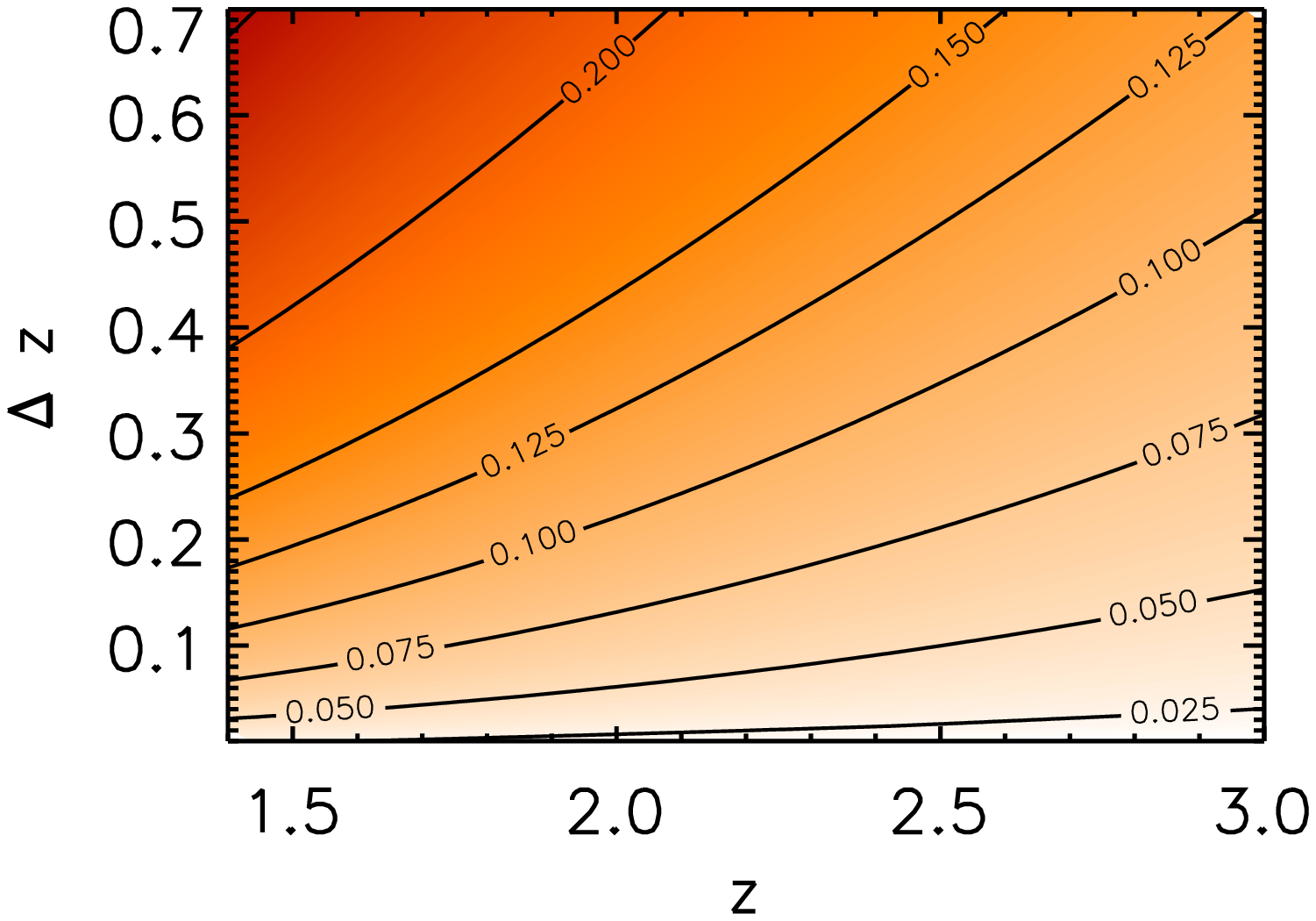}
\includegraphics[width=0.49\textwidth]{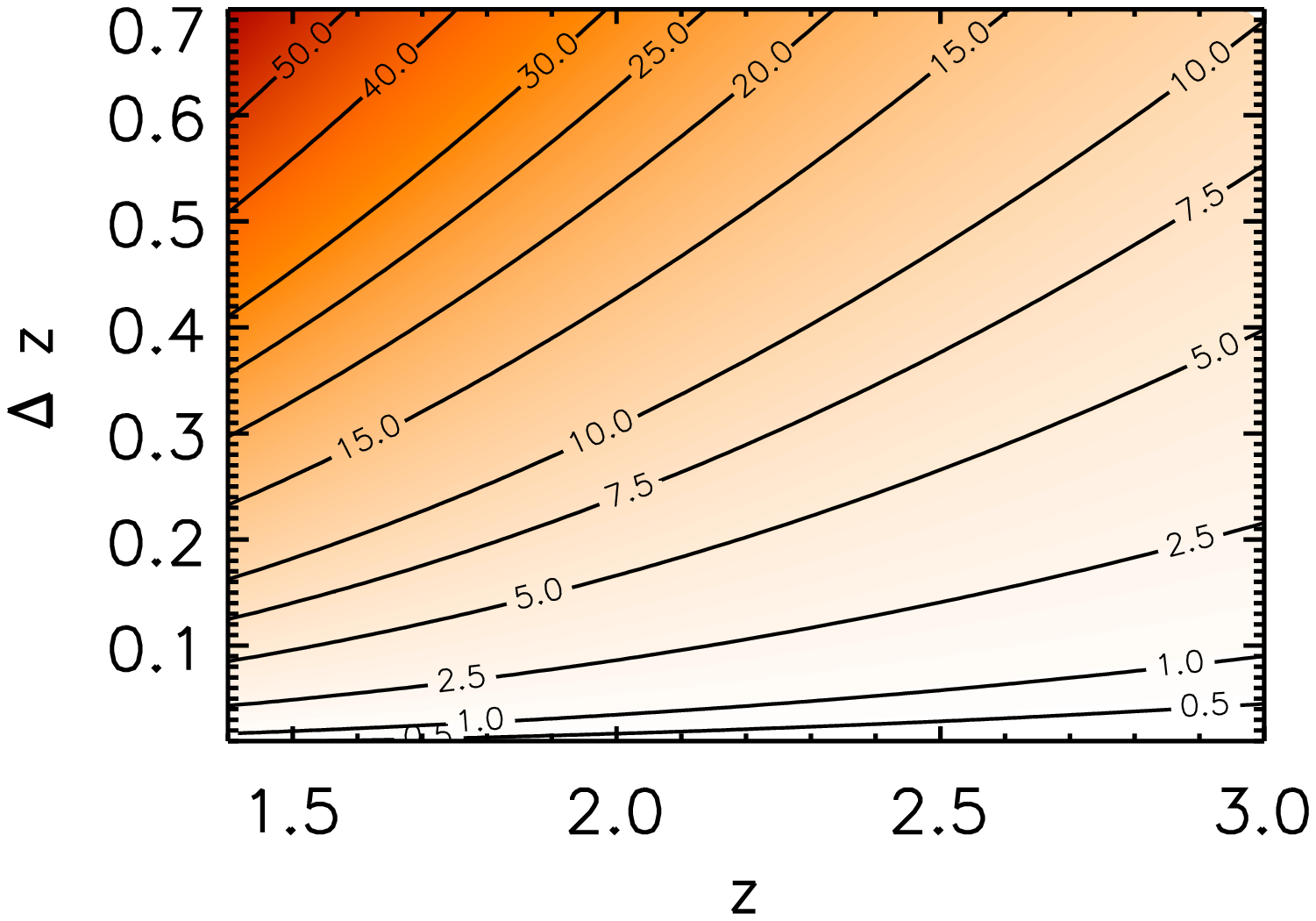}
\includegraphics[width=0.49\textwidth]{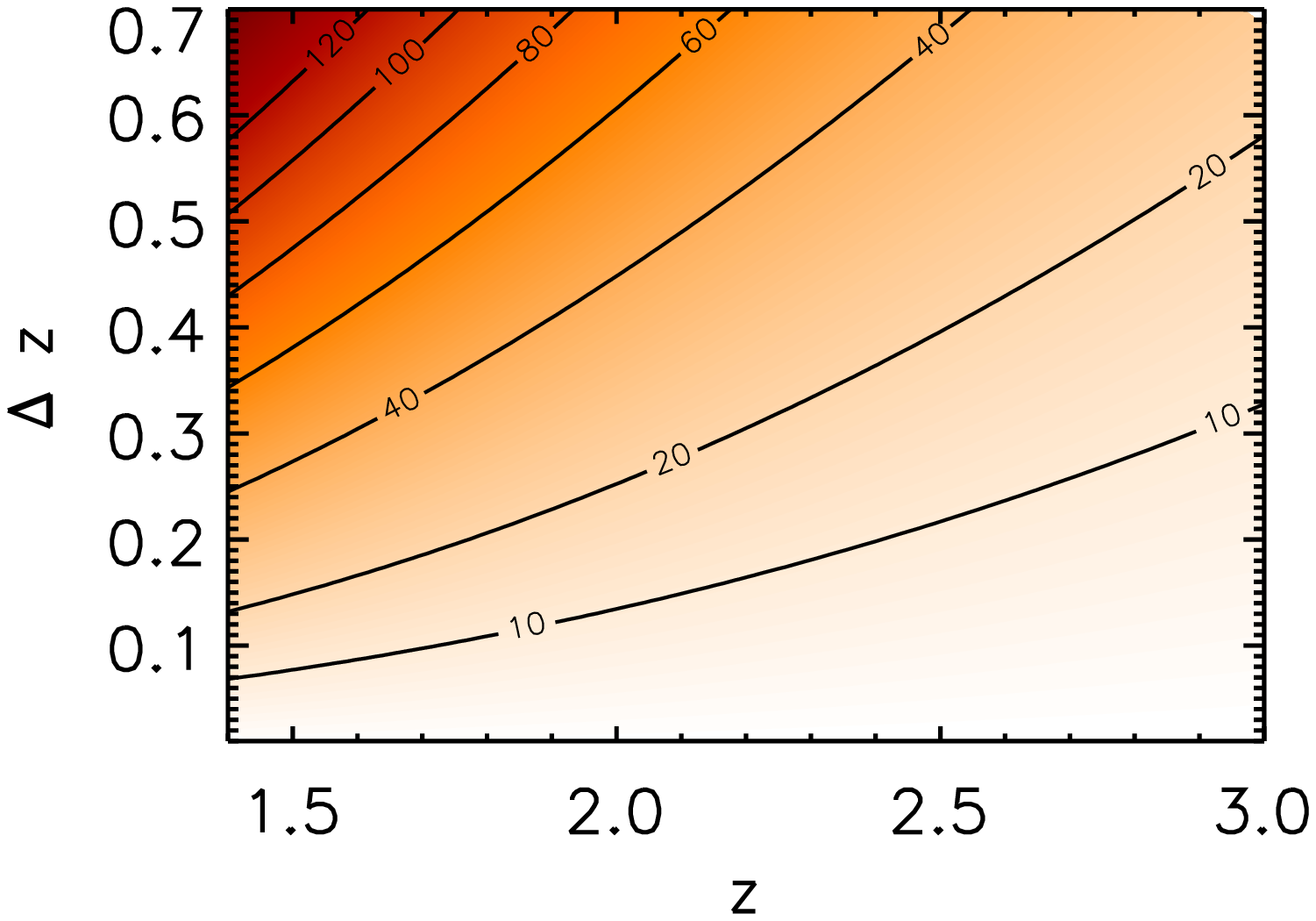}
\caption{\label{Fig:N_dFK} Estimates of the standard deviation of the correlator $
\delta\flux \kappa$ (upper panels) and $\delta\flux^2 \kappa$ (lower panels) along 
a \textit{single} line-of-sight
  as a function of the source redshift $z$ and of the length of the
  measured spectrum $\Delta z$, for convergence maps recontructed from Planck 
(left panels) and ACTPOL (right panels). As before, we set $A=\beta=1$, 
effectively turning off the physics of IGM, to make apparent the physics of structure 
formation.}
\end{figure*}

We start by considering the variance of $\delta\flux\,\kappa$. Setting $r=1$ in Eq.~
(\ref{eq:deltaF2mK2_1}) and using Wick's theorem we obtain 
\begin{eqnarray}
 \langle \delta_q\,\delta_{q'}\, \delta_c\,\delta_{c'}\rangle&\approx&
2\langle \delta_q \delta_c\rangle 
\langle \delta_{q'}\delta_{c'}\rangle+\langle \delta_q \delta_{q'}\rangle 
\langle \delta_{c}\delta_{c'}\rangle.
\end{eqnarray} 
We notice immediately that the first term is twice the square of $\langle\delta\flux\,
\kappa\rangle$, while the second term is proportional to two correlation function 
characterized by cutoffs acting \textit{either} on the modes that are parallel \textit
{or} perpendicular to the los, \textit{but not on both}. It is then possible to show that
\begin{eqnarray}
\langle \delta_{q}\delta_{q'}\rangle
&=&D(\chi_q)\,D(\chi_{q'})\left[H_0^{(0)}(\Delta\chi_q;\infty,\,k_L/\sqrt{2})\right.\vs
&-&\left. H_0^{(0)}(\Delta\chi_q;\infty,\,\bar{k}/\sqrt{2})\right],\label{Adqdq}\\
\langle \delta_{c}\delta_{c'}\rangle
&=&D(\chi_c)\,D(\chi_{c'})\,H_0^{(0)}(\Delta\chi_c;k_C/\sqrt{2}, \infty)\label{Adcdc},\\
 \langle\delta_q\delta_{c}\rangle&=&D(\chi_c)D(\chi_q)\vs
&\times&
\left[H_0^0(\Delta\chi;k_C, k_L)-H_0^0(\Delta\chi;k_C, \bar{k})\right],\\
\langle\delta_q^2\rangle&=&D^2(\chi_q)\left[\bar{H}_0^{(0)}(\chi_q,k_L)
+\bar{H}_0^{(0)}(\chi_q,\bar{k})\right],
\end{eqnarray}
where the last two equations have been added here for sake of completeness, as 
they will be useful in what follows. 
The variance of $\delta\flux \kappa$ is then
\begin{eqnarray}
 \sigma_1^2&\approx&\langle\delta\flux\,\kappa\rangle^2\vs
&+& \left(A\beta\frac{3H_0^2\Omega_m}{2c^2}\right)^2
\int_0^{\chi_F}d\chi_c \frac{W_L^2(\chi_c,\chi_F)}{a^2(\chi_c)}D^2(\chi_c)
\vs
&\times&\int_{\chi_i}^{\chi_Q}d\chi_q D^2(\chi_q)
\int_{-\Delta\chi_{c,0}}^{\Delta\chi_{c,0}} d\Delta\chi_c
H_0^{(0)}(\Delta\chi_c;k_C/\sqrt{2}, \infty)\vs
&\times&
\int_{-\Delta\chi_{q,0}}^{\Delta\chi_{q,0}} d\Delta\chi_q
\left[H_0^{(0)}(\Delta\chi_q;\infty,\,k_L/\sqrt{2})\right.\vs
&-&\left. H_0^{(0)}(\Delta\chi_q;\infty,\,\bar{k}/\sqrt{2})\right].\label{var_dK_full}
\end{eqnarray}
In the upper panels of Fig.~\ref{Fig:N_dFK} we show the values obtained for the 
\textit{standard deviation} of $\delta\flux \kappa$ for two different CMB 
experiments' resolution, again turning off the IGM physics evolution and focusing 
on the growth of structure. 

\subsubsection*{The variance of $\delta\flux^2\kappa$}

Setting $r=2$ in Eq.~(\ref{eq:deltaF2mK2_1}), we then apply Wick's theorem to $
\langle\delta^{2}_q\delta^{2}_{q'}\delta_c\delta_{c'}\rangle$. Neglecting again 
terms where the $\delta_c$'s are not contracted with one another, we obtain
\begin{eqnarray}
 \langle\delta^{2}_q\delta^{2}_{q'}\delta_c\delta_{c'}\rangle&\approx&
2\langle\delta^{2}_q\delta_c\rangle\langle\delta^{2}_{q'}\delta_{c'}\rangle\vs
&+&\langle\delta_{c}\delta_{c'}\rangle\,\left(\langle\delta^2_{q}\rangle\langle
\delta^2_{q'}\rangle+2\langle\delta_q\delta_{q'}\rangle^2\right),\label
{N:d2qd2qdcdc}
\end{eqnarray} 
which then leads to the expression for $\sigma_2^2$
\begin{align}
 \sigma_2^2 &\approx \langle\delta\flux^2\,\kappa\rangle^2\vs
&+\left(A\beta\frac{3H_0^2\Omega_m}{2c^2}\right)^2\int_0^{\chi_F}d\chi_c \frac
{W_L^2(\chi_c,\chi_F)}{a^2(\chi_c)}D^2(\chi_c)\vs
&\times\int_{-\Delta\chi_{c,0}}^{\Delta\chi_{c,0}} d\Delta\chi_c
H_0^{(0)}(\Delta\chi_c;k_C/\sqrt{2}, \infty)\vs
&\times \left\{ \left[\bar{H}_0^{(0)}(\chi_q,k_L)
+\bar{H}_0^{(0)}(\chi_q,\bar{k})\right]^2\left[\int_{\chi_i}^{\chi_Q}d\chi_q D^2
(\chi_q)\right]^2\right.\vs
&+2\int_{\chi_i}^{\chi_Q}d\chi_q D^4(\chi_q)
\int_{-\Delta\chi_{q,0}}^{\Delta\chi_{q,0}} d\Delta\chi_q
\left[H_0^{(0)}(\Delta\chi_q;\infty,\,k_L/\sqrt{2})\right.\vs
&\left.\left.- H_0^{(0)}(\Delta\chi_q;\infty,\,\bar{k}/\sqrt{2})\right]^2
\right\}.\label{eq:sigma_2^2}
\end{align}

In the lower panels of Fig.~\ref{Fig:N_dFK} we show the estimates for
the \textit{standard deviation} $\delta\flux^2 \kappa$ along a
\textit{single} line-of-sight for the two different CMB experiment.
We note in Fig.~\ref{Fig:N_dFK} the same trends that have been pointed
out for the correlator itself in Fig.~\ref{Fig:dK_S_3_2} and
\ref{Fig:S_3_2}: the standard deviation of $\delta\flux \kappa$ and of
$\delta\flux^2 \kappa$ increase almost linearly with increasing length
of the \Lya spectrum $\Delta z$ and it decreases as the source
redshift $z$ is increased because of the fact that the spectrum probes
regions that are less clumpy. Also,
by increasing the resolution of the CMB experiment used to reconstruct
the convergence map, the deviation of $\delta\flux \kappa$ and
$\delta\flux^2 \kappa$ also increase: if on one hand more modes carry
more information, on the other hand they also carry more cosmic
variance.

One last aspect to note here is that while the signal for
$\langle\delta\flux^2\kappa\rangle$ arises from a three point
correlation function (which in the gaussian approximation would yield
zero), the dominant terms contributing to its variance arise from
products of two point correlation functions. In particular, it is
possible to show that the terms appearing in the second line of
Eq.~(\ref{N:d2qd2qdcdc}) significantly outweight the square of the signal
that appears in the first line.

\begin{figure*}
\includegraphics[width=0.49\textwidth]{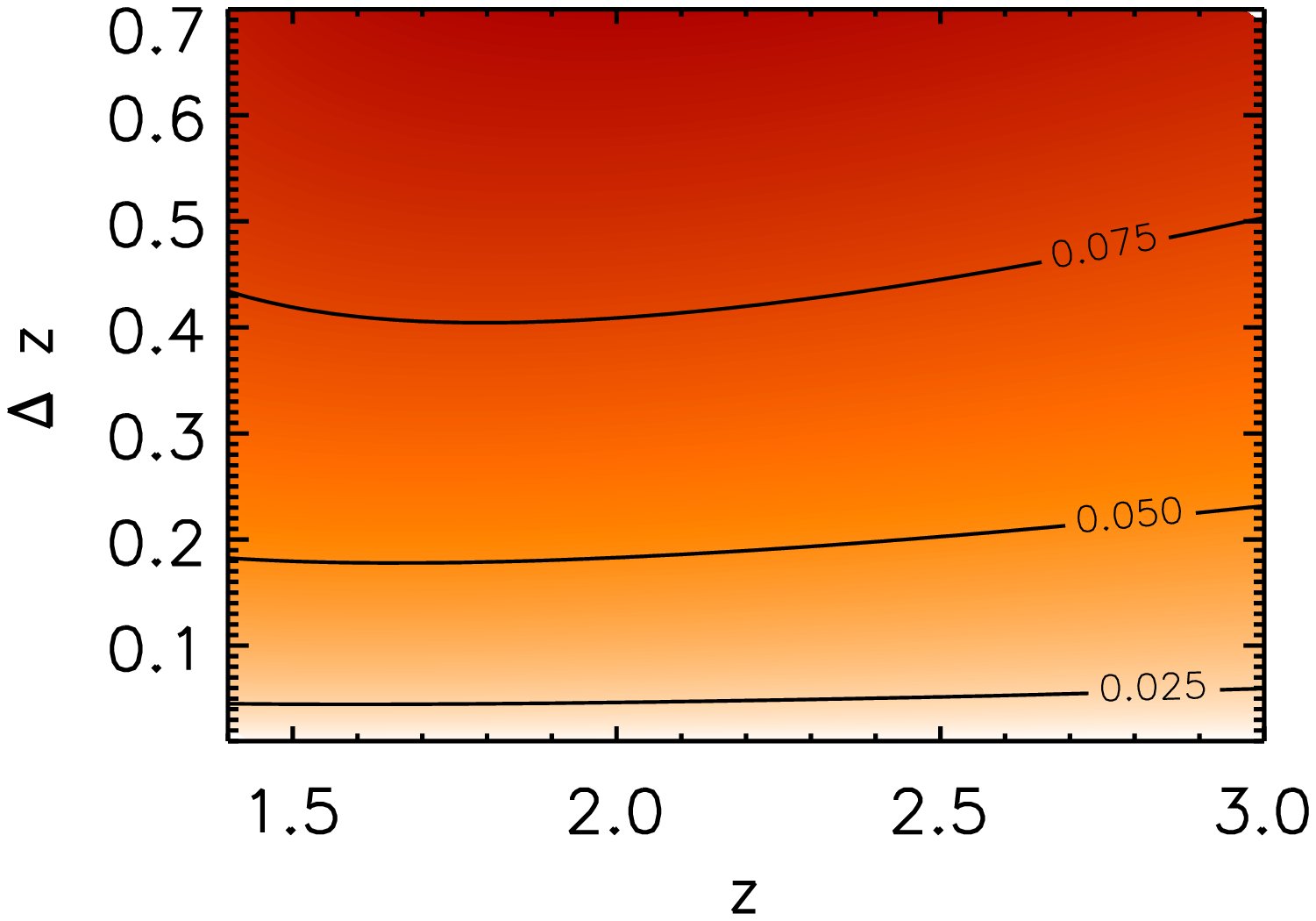}
\includegraphics[width=0.49\textwidth]{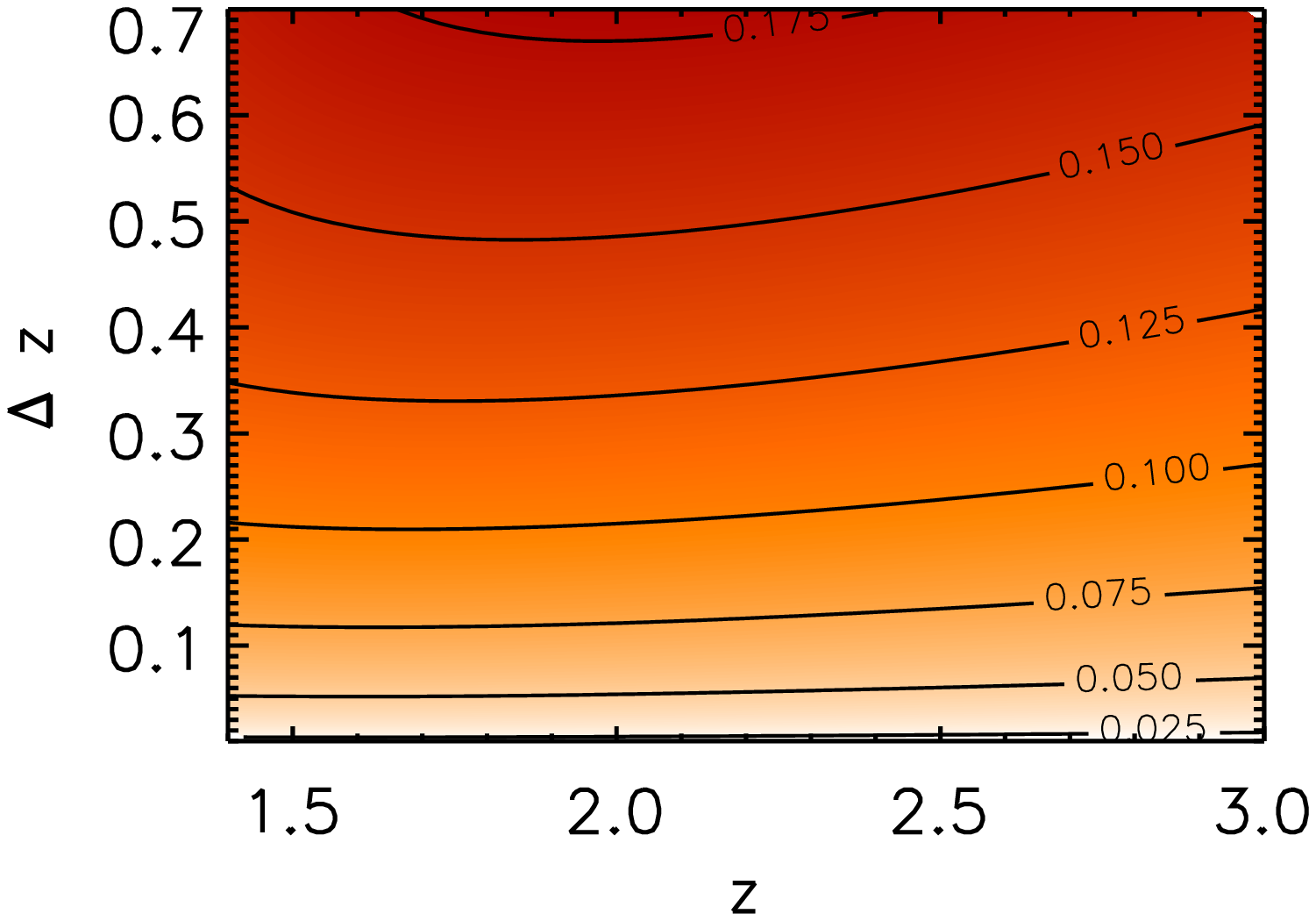}
\includegraphics[width=0.49\textwidth]{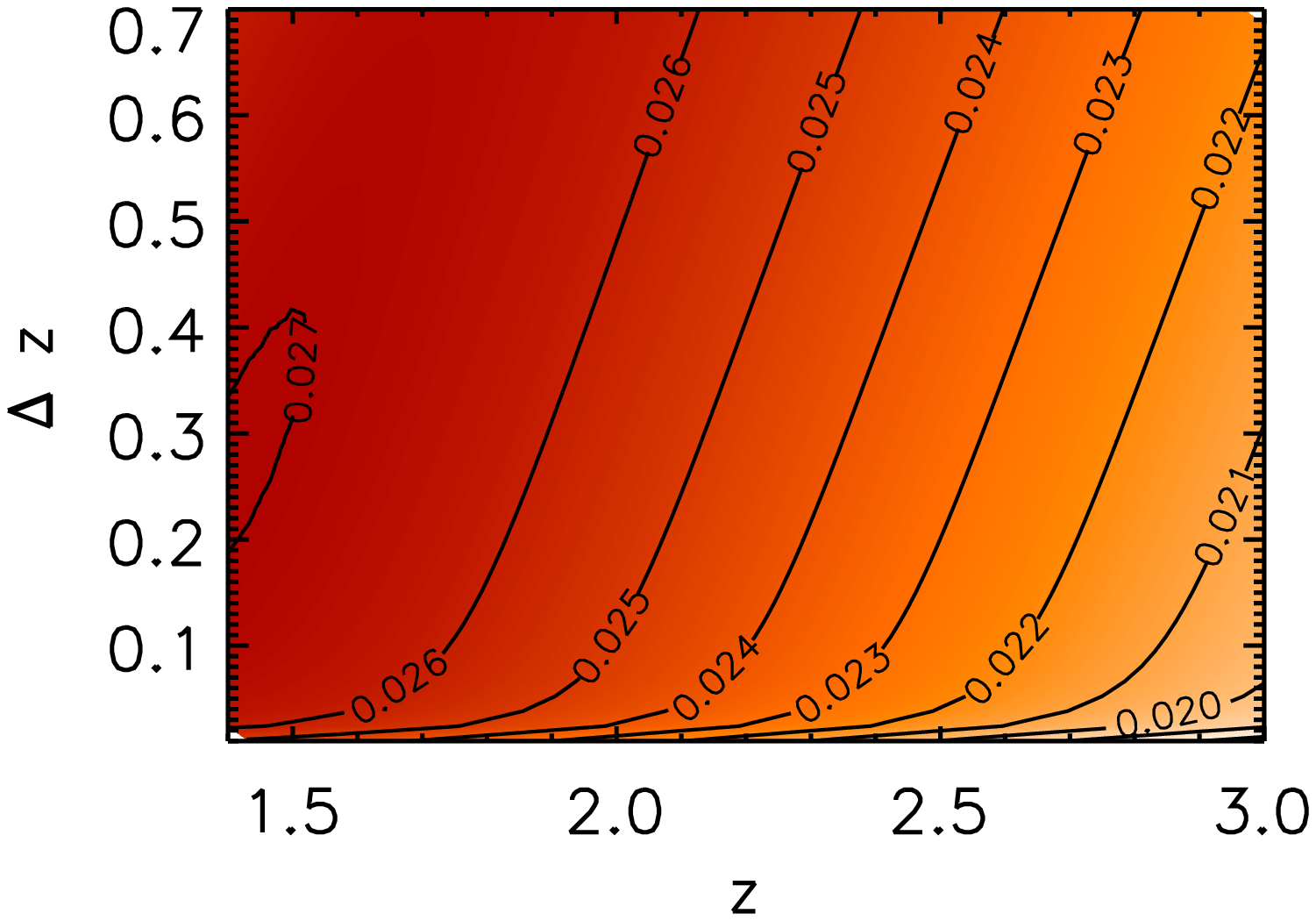}
\includegraphics[width=0.49\textwidth]{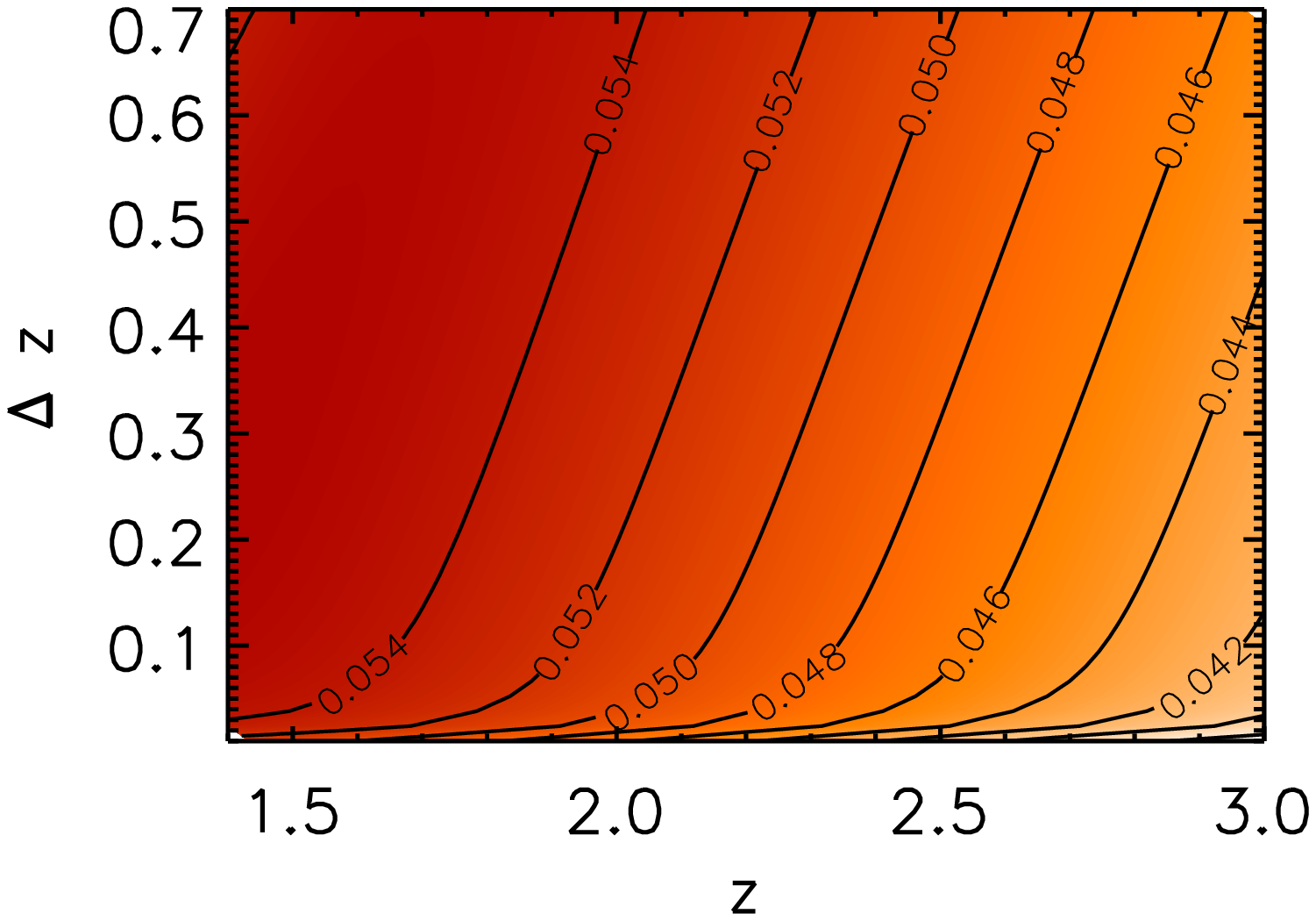}
\caption{\label{Fig:dk_SN_3_2} Estimates for the signal-to-noise
  ratios for the observation of the correlators $\langle \delta\flux
  \kappa\rangle$ along a \textit{single} line-of-sight as a function
  of the source redshift $z$ and of the length of the measured
  spectrum $\Delta z$, for Planck (left panels) and ACTPOL (right
  panels). As long as the functions $A$ and $\beta$ can be assumed to
  be constant in the redshift range spanned by the \Lya spectrum,
  these result do not depend on the specific value taken by the
  latter.}
\end{figure*}

\subsection{Signal-to-Noise ratio}
\label{sn}
We now have all the pieces to assess to what extent the
$\langle\delta\flux^r\kappa\rangle$ correlations will be detectable by
future observational programs. Even before moving to plot the S/N
ratios for $\delta\flux\kappa$ and $\delta\flux^2\kappa$ it is
possible to point out a couple of features of these ratios. First, we
note that the S/N ratio for $\delta\flux\kappa$ and
$\delta\flux^2\kappa$ do present a radical difference in their
dependence on the QSO source redshift. This is because the signal for
$\delta\flux^2\kappa$ is characterized by mode coupling, whereas the
dominant contributions to the variance are not. Physically, the
signal for $\delta\flux^2\kappa$ is more sensitive to the growth of
structure with respect to its variance: while for the former the
growth of long wavelength modes enhances the growth of structure on
small scales, for the latter long and short wavelength modes grow independently at
the same rate. Mathematically, this is apparent when comparing
Eq.~(\ref{eq:deltaFmK_2}) with Eq.~(\ref{eq:sigma_2^2}): while the
$\lg\delta\flux^2\kappa\rg$ signal carries four powers of the growth factor, the
dominant terms contributing to its variance carry only six. In this
case then the S/N is characterized by four growth factors in the
numerator and only three in the denominator, thus leading to a
``linear'' dependence of S/N on the redshift (modulo integration over
the los and behaviour of the lensing window function). Note that this
is in stark contrast with the $\langle\delta\flux\kappa\rangle$ case,
where the signal is not characterized by mode coupling and the number
of growth factors are equal for the signal and its standard deviation,
thus leading to a S/N ratio with no dependence on the source's
redshift.

Second, we note that S/N does not depend on the value of any
constant. In particular, regardless of their redshift dependence, the
S/N ratio will not depend on the functions $A$ and $\beta$ used to
describe the IGM. This is of course very important since in such a
way, at least in linear theory and using the FGPA at first order, the
dependence on the physics of the IGM cancels out when computing the
S/N ratio.

In Fig.~\ref{Fig:dk_SN_3_2} we show the estimates for the S/N
\textit{per los} of the $\lg\delta\flux\kappa\rg$ (upper panels) and
$\lg\delta\flux^2\kappa\rg$ (lower panels) measurements. As expected,
while the S/N for $\lg\delta\flux\kappa\rg$ does not show any strong
redshift dependence, the S/N for $\lg\delta\flux^2\kappa\rg$ decreases
linearly with increasing source redshift: the growth of structure is
indeed playing a role and shows that QSOs lying at lower redshift will
yield a larger S/N. Also, in both cases an increase in the resolution
of the experiment measuring the convergence field translates in a
larger S/N and in a larger derivative of the S/N with respect to
$\Delta z$. This is not surprising, as it is reasonable to expect that
a higher resolution convergence map will be carrying a larger amount
of information about the density field.

All this suggests that depending on what is the correlator that one is
interested in measuring, different strategies should be pursued. In
case of $\lg\delta\flux \kappa\rg$ increasing the length of the
spectra will provide a better S/N. In case of
$\lg\delta\flux^2\kappa\rg$, however, Fig.~(\ref{Fig:dk_SN_3_2})
suggests that an increase in the number of quasar will be more
effective in producing a large S/N, whereas an increase in the
redshift range spanned by the spectrum will increase the S/N only
marginally.

Having obtained the S/N per los, we can then estimate the total S/N
that will be obtained by cross-correlating the BOSS sample
($1.6\cdot10^5$ QSOs) and the proposed BigBOSS sample
\cite{Schlegel:2009uw} ($10^6$ QSOs) with the convergence map measured
by Planck or by the proposed ACTPOL experiment
considered. Assuming a mean QSO redshift of $\bar{z}=2.5$ and a mean
\Lya spectrum length of $\Delta z=0.5$, a rough estimate of the S/N for the 
measurements
of $\langle \delta\flux\kappa \rangle$ and of $\langle
\delta\flux^2\kappa \rangle$ are given in Tab.~\ref{Tab1:SN_dFK} and
\ref{Tab1:SN_dF2K}.
\begin{table}
\begin{center}
\begin{tabular}{cccc}
\hline
CMB Exp. & S/N  & Total S/N & Total S/N\\
	 & per \textit{los} & in BOSS & in BigBOSS\\
\hline
\hline
Planck & 0.075 &  30 & 75\\
ACTPOL & 0.130 & 52 & 130\\
\hline
\end{tabular}
\end{center}
\caption{\label{Tab1:SN_dFK} Estimates of the total and per single \textit{los}
  signal--to--noise (S/N) of the $\langle\delta\flux\,\kappa\rangle$
  cross--correlation for different CMB experiments combined with BOSS
  and BigBOSS.}
\label{table1}

\end{table}%

It is necessary to point out here that despite that the value of the
S/N for $\langle \delta\flux\kappa \rangle$ is almost three times
larger than the one for $\langle \delta\flux^2\kappa \rangle$, the
actual measurement of the former correlator strongly depends on the
ability of fitting the continuum of the \Lya spectrum. The $\langle
\delta\flux^2\kappa \rangle$ correlator, on the other hand, is
sensitive to the interplay between long and short wavelength modes and
as such should be less sensitive to the continuum fitting
procedure. Therefore, even if it is characterized by a lower S/N, it
may actually be the easier to measure in practice.  The numbers
obtained above are particularly encouraging since the S/N values are
typically very large and well above unity. 

\begin{table}
\begin{center}
\begin{tabular}{cccc}
\hline
CMB Exp. & S/N  & Total S/N & Total S/N\\
	 & per \textit{los} & in BOSS & in BigBOSS\\
\hline
\hline
Planck & 0.024 &  9.6 & 24\\
ACTPOL & 0.05 & 20.0 & 50\\
\hline
\end{tabular}
\end{center}
\caption{\label{Tab1:SN_dF2K} Estimates of the total and per single \textit{los}
  signal--to--noise (S/N) of the $\langle\delta\flux^2\,\kappa\rangle$
  cross--correlation for different CMB experiments combined with BOSS
  and BigBOSS.}
\label{table2}
\end{table}%

\subsection{Analysis}
\label{spectral}
Having developed a calculation framework for estimating $\langle
\delta\flux^r\kappa \rangle$ and the S/N for their measurement, we
turn to estimate what is the range of \Lya wavelengths contributing to
the signal and what is the effect of changing the parameters that
control the experiments' resolution.

\subsubsection*{Spectral Analysis}

We investigate here how the different \Lya modes contribute to the
correlators. This should tell us whether long wavelength modes have
any appreciable effect on our observables and what is the impact of
short and very short wavelength modes (in particular the ones that are
expected to have entered the non-linear regime).

Since the mean flux $\bar{\flux}$ appearing in the definition of the
flux fluctuation $\delta\flux=(\flux-\bar{\flux})/\bar{\flux}$ is a
\textit{global} quantity which is usually estimated from a
statistically significant sample of high resolution QSO spectra (see
the discussion in \cite{seljak03} for the impact that such quantity
has on some derived cosmological parameters), $\delta\flux$ is
sensitive also to modes with wavelengths longer than the \Lya
spectrum. These modes appear as a ``background'' in each spectra but
they still have to be accounted for when crosscorrelating
$\delta\flux$ with $\kappa$ because the fluctuation in the flux is
affected by them. More specifically, a QSO that is sitting in an
overdense region that extends beyond the redshift range spanned by its
spectrum will see its flux decremented by a factor that in its
spectrum will appear as \textit{constant} decrement. On the other
hand, if the QSO spectrum extends beyond the edge of such overdensity,
this mode would appear as a fluctuation (and not as a
background) in the spectrum. This extreme scenario is somewhat
mitigated by the fact that present and future QSO surveys will have
many QSOs with los separated by few comoving Mpc \cite{BOSS_th}: as such,
fluxes from neighboring QSO lying in large overdense regions should
present similarities that should in principle allow to detect such
large overdensities in 3D tomographical studies \cite{saitta08}.

\begin{figure}[b]
\includegraphics[width=0.49\textwidth]{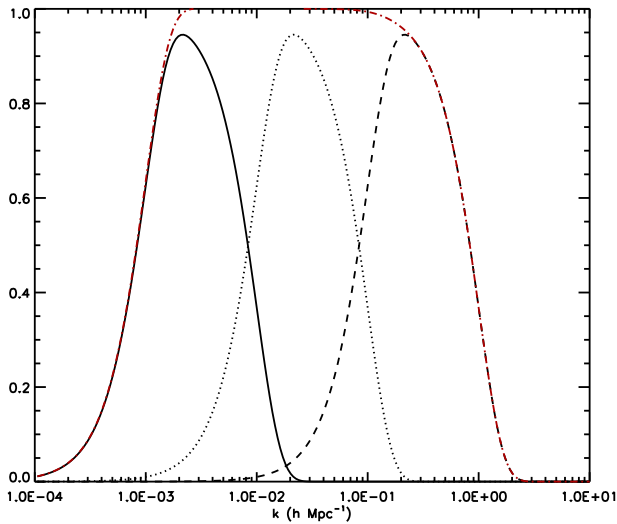}
\caption{\label{Fig:Filters} Three filters used to calculate the
  contribution of the different modes to the correlators, their
  variance and the SN ratio. The filters have $\{k_l=10^{-3},
  k_L=10^{-2}\}$ (solid curve), $\{k_l=10^{-2}, k_L=10^{-1}\}$ (dotted
  curve) and $\{k_l=10^{-1}, k_L=1\}$ (dashed curve). Also shown is
  the sum of the filters (red dashed-dotted curve).}
\end{figure}

To measure the contributions of the different modes to the
correlators, we vary $k_l$ and $k_L$ to build appropriate filters. As
can be seen from Fig.~\ref{Fig:Filters}, where three such filters are
plotted for $\{k_l=0.001, k_L=0.01\}$, $\{k_l=0.01, k_L=0.1\}$ and
$\{k_l=0.1, k_L=1\}$, the gaussian functional form assumed for the
window function does not provide very sharp filters (hence this
spectral analysis will not reach high resolution). 
Also, if $k_L=10\,k_l$ then the filters add exactly to
one. This allows us to measure the contributions of the different
wavenumber decades to the correlators and its standard deviation.

\begin{table}
\begin{tabular}{ccccc}
\hline
$k_l$ & $k_L$ & $|\langle\delta\flux\kappa\rangle|$ & $\sigma_{\delta\flux\kappa}
$ & Ratio \\
\hline\hline
1.00e-04 & 1.00e-03 & 1.66e-04 & 1.77e-04 & 9.39e-01\\
1.00e-03 & 1.00e-02 & 1.20e-03 & 1.21e-03 & 9.87e-01\\
1.00e-02 & 1.00e-01 & 2.12e-04 & 6.29e-04 & 3.37e-01\\
1.00e-01 & 1.00e+00 & 6.11e-07 & 1.42e-03 & 4.30e-04\\
1.00e+00 & 1.00e+01 & 7.26e-08 & 2.44e-03 & 2.97e-06\\
\hline
\end{tabular} 
\caption{\label{Table:dK_Spectral} Contribution of the different
  wavenumbers (split over decades) to the absolute value of the correlator
  $\langle\delta\flux\kappa\rangle$, its standard deviation
  $\sigma_{\delta\flux\kappa}$ and ratio of the two quantities. In this calculation we 
took into account the evolution of $A$ with redshift.}
\end{table}

Table \ref{Table:dK_Spectral} and \ref{Table:Spectral} summarize the
results for $\langle\delta\flux\kappa\rangle$ and
$\langle\delta\flux^2\kappa\rangle$ respectively. Considering
$\langle\delta\flux\kappa\rangle$ we note immediately that the signal
and the S/N ratio both peaks around $k\simeq 10^{-2}$ $\hMpcI$, as
expected from the fact that this signal is proportional to the two
point correlation function, which in turn receives its largest
contribution from the wavelengths that dominate the power
spectrum: isolating the long wavelength modes of the \Lya flux would allow to
increase the S/N. However, this procedure is sensibly complicated by
the continuum fitting procedures that are needed to correctly
reproduce the long wavelength fluctuations of the \Lya flux. The behavior of the 
variance is interesting, as in the first
three decades shows an oscillating behavior. This is due to
the different weights of the two terms appearing in
Eq.~(\ref{var_dK_full}) for each range of wavelengths. In particular,
for $k\lesssim 10^{-2}$ $\hMpcI$ the variance of
$\langle\delta\flux\kappa\rangle$ is dominated by the first term, that
is just the square of the signal. However, as the signal gets smaller
with increasing $k$, for $k\gtrsim 10^{-1}$ $\hMpcI$ it is the second term that
dominates the variance. 

\begin{table}
\begin{tabular}{ccccc}
\hline
$k_l$ & $k_L$ & $\langle\delta\flux^2\kappa\rangle$ & $\sigma_{\delta
\flux^2\kappa}$ & Ratio \\
\hline\hline
1.00e-04 & 1.00e-03 & 1.08e-04 & 2.18e-02 & 4.99e-03\\
1.00e-03 & 1.00e-02 & 6.69e-03 & 1.96e-01 & 3.40e-02\\
1.00e-02 & 1.00e-01 & 5.92e-02 & 1.31e+00 & 4.52e-02\\
1.00e-01 & 1.00e+00 & 3.39e-01 & 7.06e+00 & 4.80e-02\\
1.00e+00 & 1.00e+01 & 9.92e-01 & 2.07e+01 & 4.79e-02\\
\hline
\end{tabular} 
\caption{\label{Table:Spectral} Contribution of the different
  wavenumbers (split over decades) to the correlator
  $\langle\delta\flux^2\kappa\rangle$, its standard deviation
  $\sigma_{\delta\flux^2\kappa}$ and ratio of the two quantities. In this calculation 
we took into account the evolution of $A$ with redshift.}
\end{table}

\begin{figure*}
\includegraphics[width=0.49\textwidth]{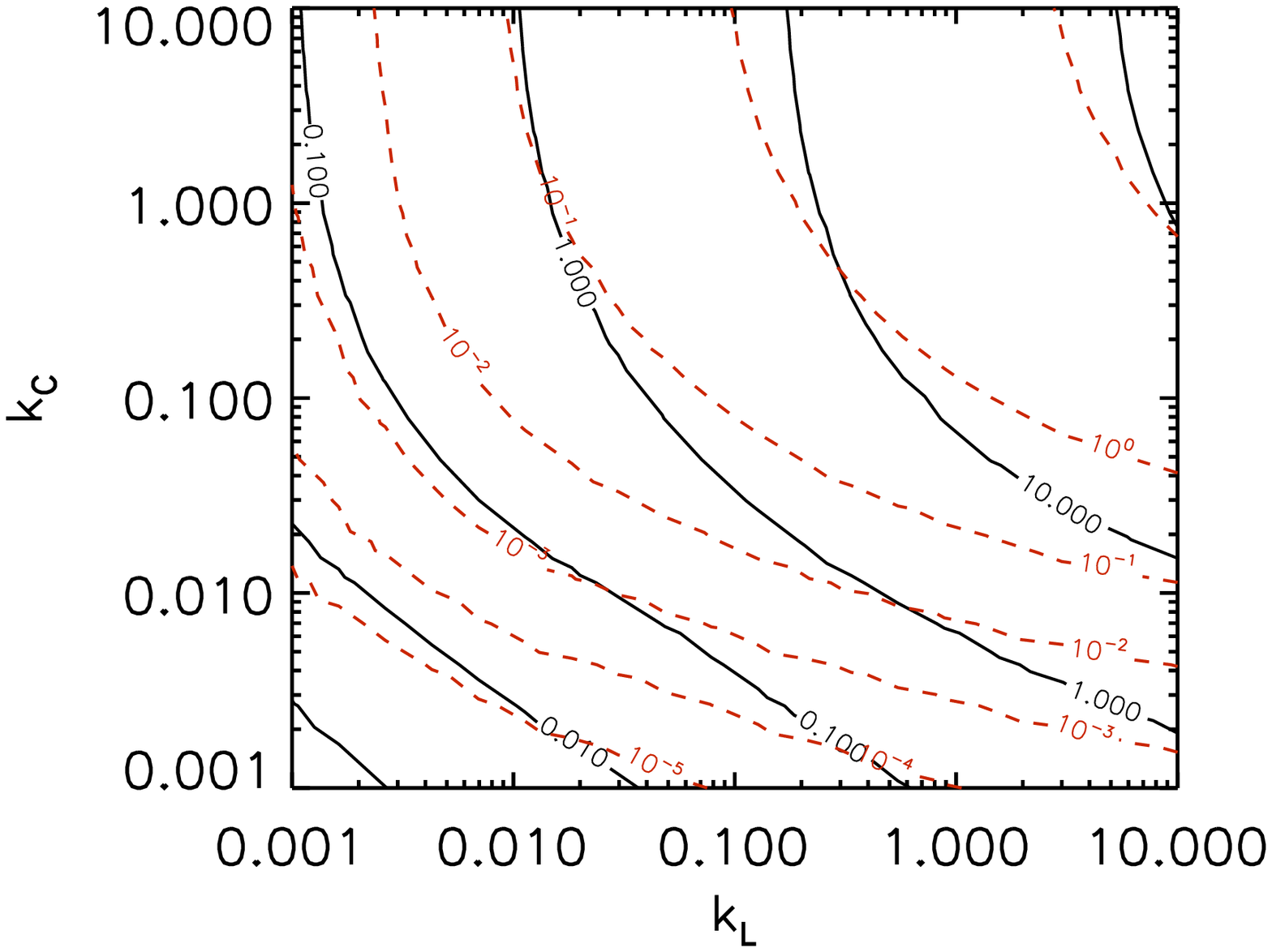}
\includegraphics[width=0.49\textwidth]{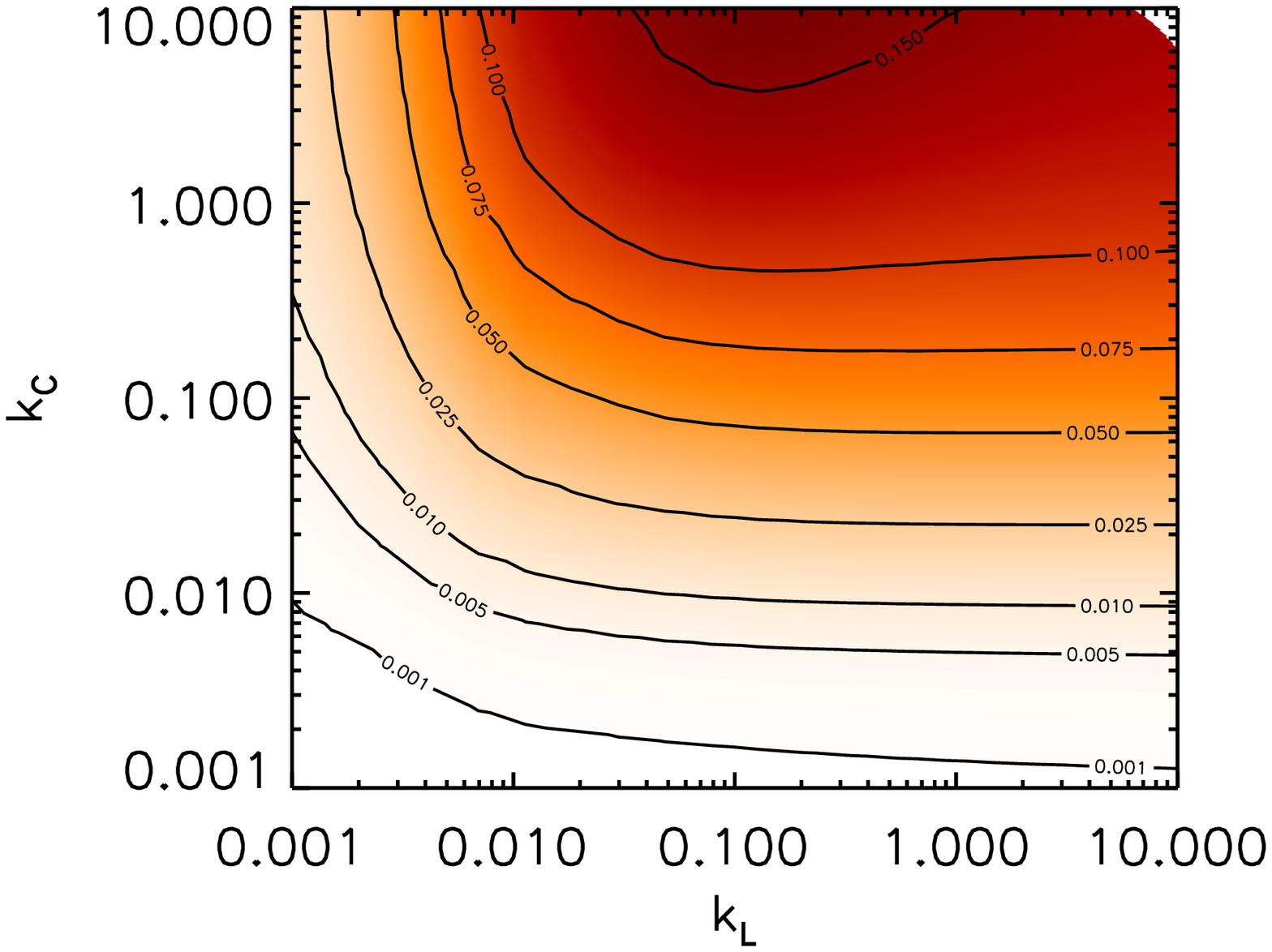}
\caption{\label{Fig:kL_kC} Value of $\lg\delta\flux^2 \kappa\rg$ (left
  panel, red dashed contours), of its standard deviation (left panel,
  black solid contour) and of its S/N ratio (right) for a single QSO
  lying at $z=2.6$ and whose spectrum covers $\Delta z=0.5$. Here we
  assume $k_l=0$ and $A=\beta=1$.}
\end{figure*}

Regarding $\langle\delta\flux^2\kappa\rangle$, it is necessary to
point out two aspects. First, short wavelengths (high-$k$) modes
provide the larger contribution to \textit{both} the correlator
\textit{and} its standard deviation. Second, for $k\gtrsim 10^{-2}$
$\hMpcI$ the ratio of the contribution to the correlator and to its
standard deviation remain almost constant. This means that above
$10^{-2}$ $\hMpcI$ the different frequency ranges contribute roughly
in the same proportion. This fact is both good news and bad news at the same
time. It is bad news because it means that increasing the resolution
of the \Lya spectra does not automatically translate into increasing
the \textit{precision} with which the correlator will be measured, as
the high-$k$ modes that are introduced will boost both the correlator
and its variance in the same way. On the other, this appears also to
be good news because it tells us that low resolution spectra
\textit{which do not record non-linearities on small scales} can be
successfully used to measure this correlation. To increase the S/N
ratio and to achieve a better precision for this measurement it is
better to increase the number of QSO spectra than to increase the
resolution of each single spectra. Finally, cutting off the
long-wavelength modes with $k\lesssim 10^{-2}\, \hMpcI$ should \textit{not}
have a great impact on the S/N ratio or on the measured value of the
correlator: if on one hand the contribution of the modes with $k\lesssim
10^{-2} \, \hMpcI$ are noisier due to cosmic variance, on the other
hand the absolute value of such contributions to the correlator and to
its variance are negligible compared to the ones arising from $k\gtrsim
10^{-2} \, \hMpcI$. We can see this fact also comparing the last
column of Tab.~\ref{Table:Spectral} with the right panel of
Fig.~\ref{Fig:kL_kC} where the \textit{absolute value} of the S/N
ratio is plotted for varying values of the cutoffs $k_L$ and $k_C$. By
looking at the last column of Tab.~\ref{Table:Spectral} we see that
the ratio between the correlator and its standard deviation increases
until about $k\simeq 10^{-2}\hMpcI$ where it levels off. Looking at
the right panel of Fig.~\ref{Fig:kL_kC} we notice exactly the same
trend: increasing the resolution of the spectrum $k_L$ above
$10^{-2}\hMpcI$ does not improve the dramatically the S/N ratio. This
is because from that point on each new mode contributes in almost the
same amount to the correlator \textit{and to its standard deviation}.

\subsubsection*{Dependence on experimental resolutions}

To analyze the impact of a change in the resolution of the experiments
measuring the CMB convergence map or the \Lya flux we consider a
single QSO at redshift $z_0=2.6$ whose spectrum covers $\Delta z=0.5$
and vary $k_L$ and $k_C$. In this case we set $k_l=0$.

In Fig.~\ref{Fig:kL_kC} we show the value of $\lg\delta\flux^2
\kappa\rg$, of its standard deviation and of its S/N ratio for varying
values of $k_L$ and $k_C$. We note that both the correlator and
its standard deviation increase with increasing resolution: this makes
physical sense as increasing the resolution increases both the amount
of information carried by each experiment and the cosmic variance
associated with it. Except for very low values of $k_C$, an increase
in the resolution of the \Lya spectrum is characterized by an almost
equal amount of increase in both the correlator and its cosmic
variance. This implies that the S/N becomes roughly constant for
$k_L\gtrsim 10^{-2}$ $\hMpcI$. On the other hand, increasing $k_C$
increases both the correlator and its cosmic variance only up to the
point where $k_C\simeq k_L$.

\section{Cosmological Applications}

\subsection{Neutrinos}
\label{neutrinos}

\begin{figure*}
\includegraphics[width=0.49\textwidth]{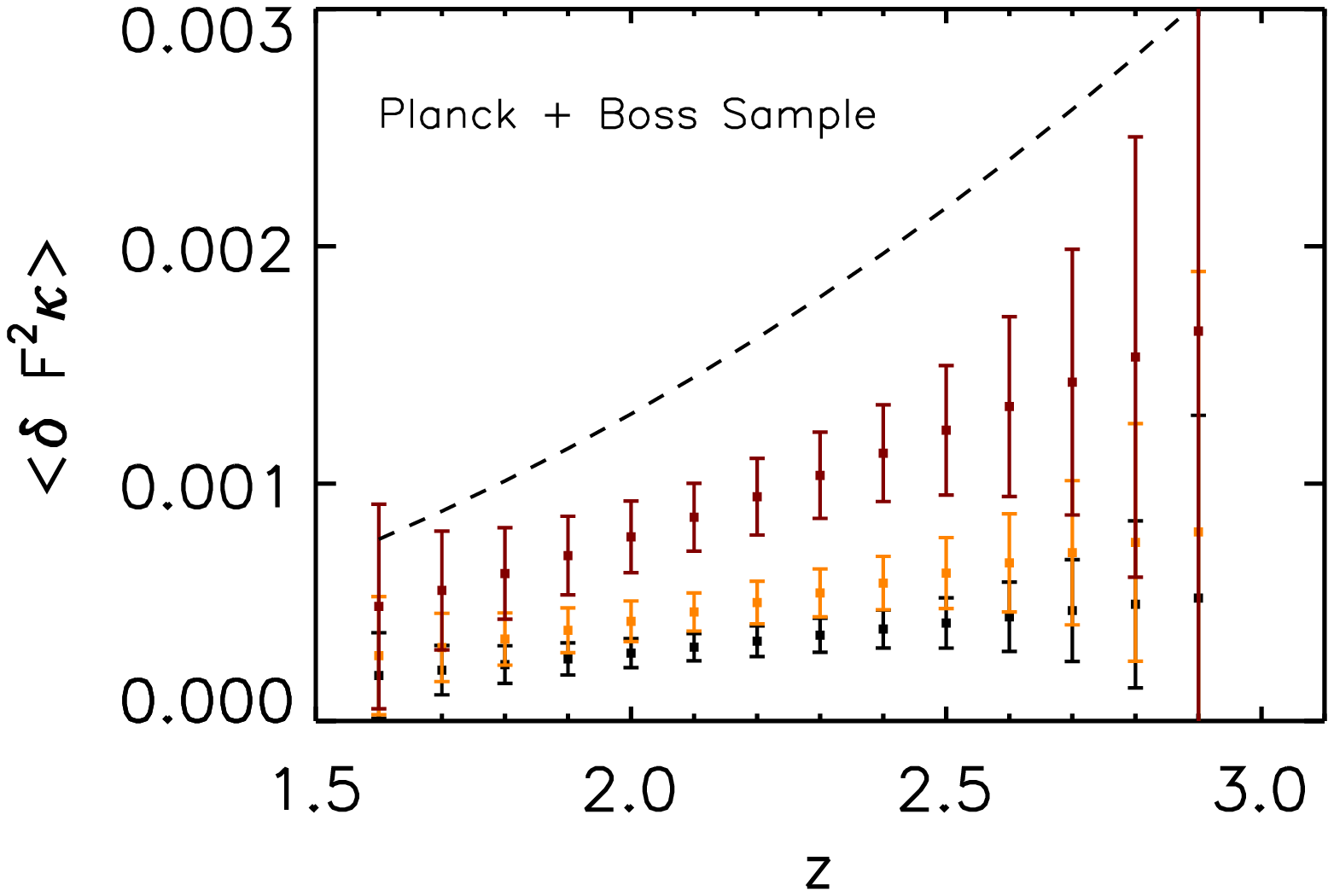}
\includegraphics[width=0.49\textwidth]{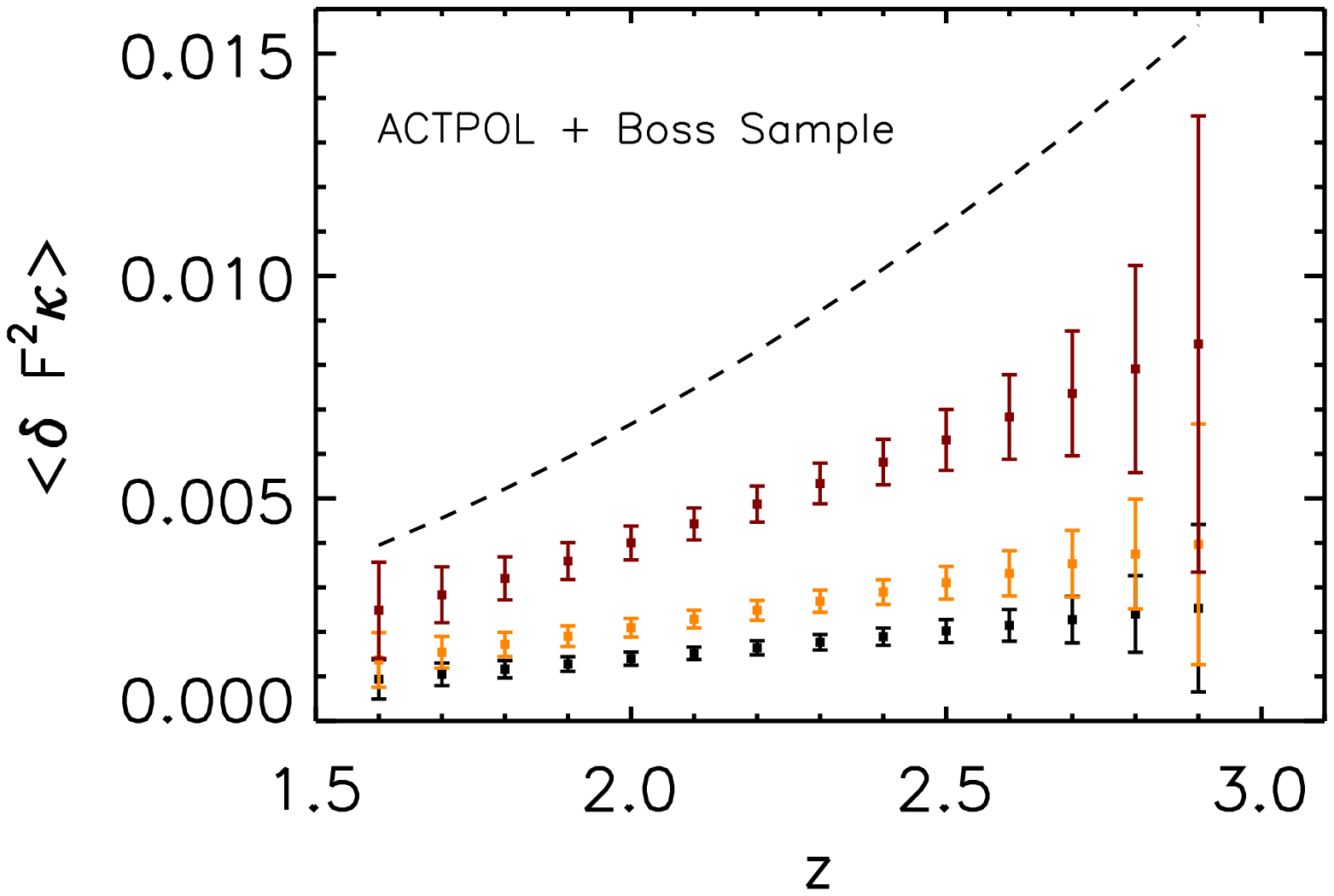}
\includegraphics[width=0.49\textwidth]{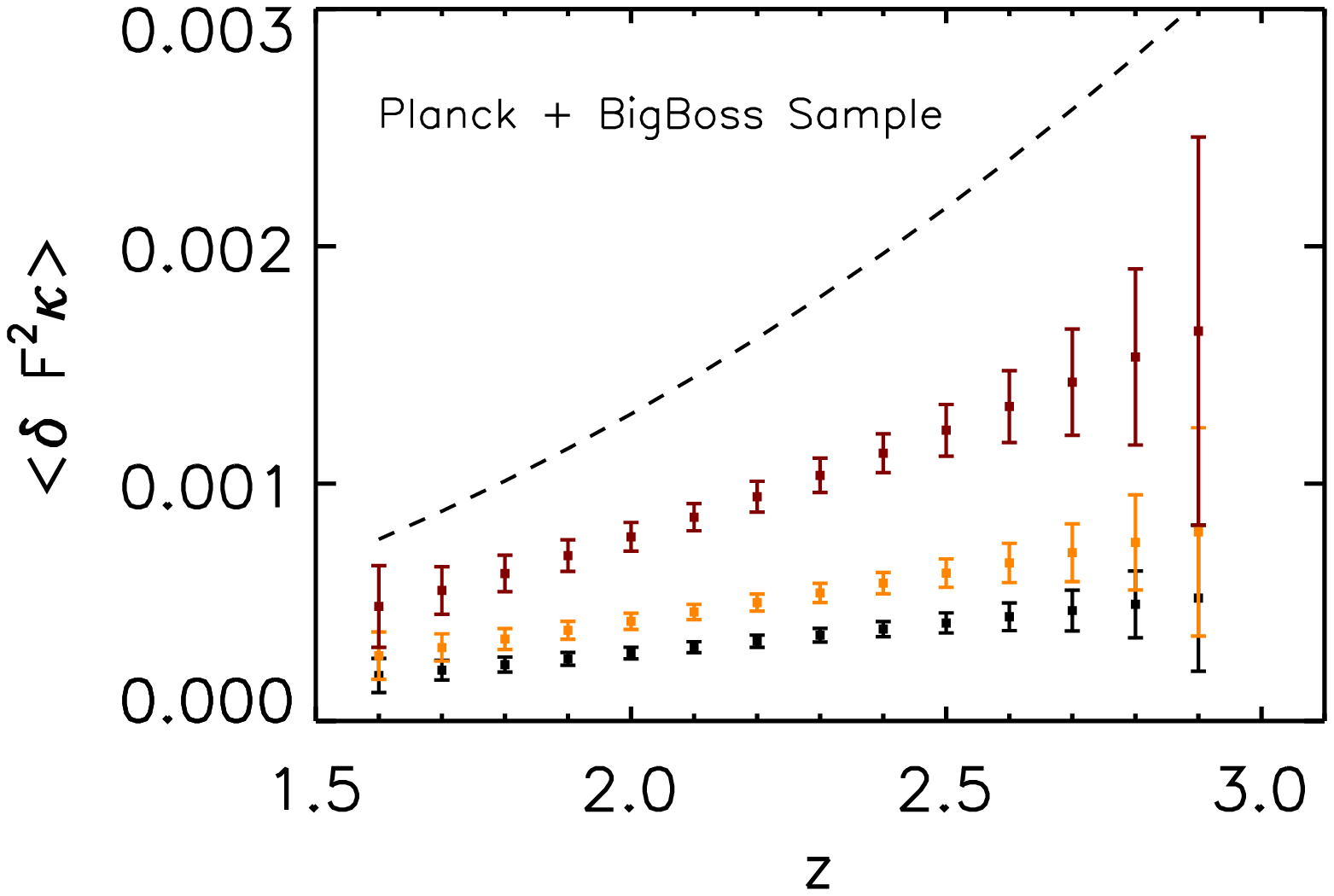}
\includegraphics[width=0.49\textwidth]{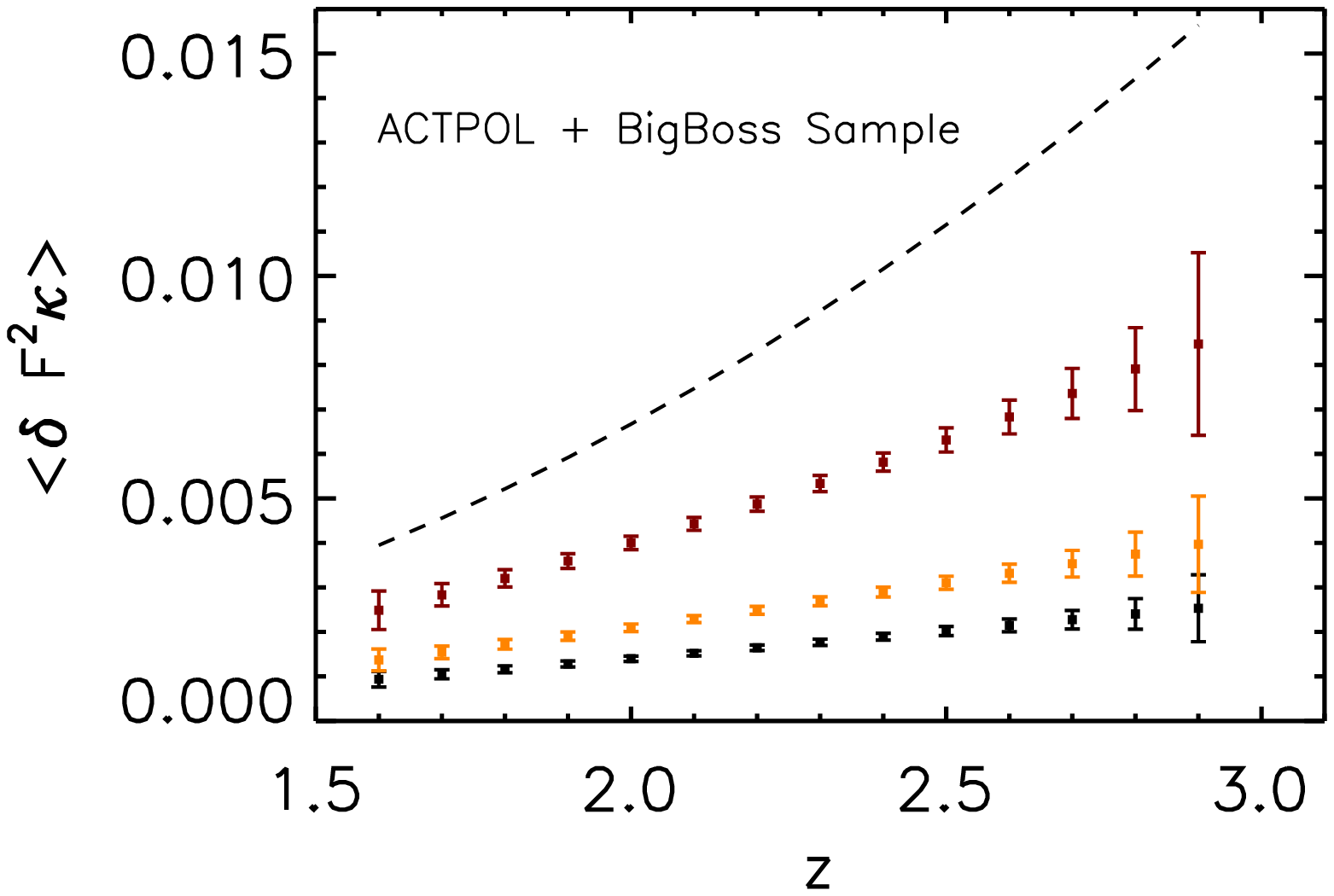}
\caption{\label{Fig:Neutrinos} Cross-correlation of the variance of
  the \Lya flux and CMB convergence as a function of redshift for the
  three different cosmological models with massive neutrinos shown in
  Table IV. Black, orange and red (with $1\sigma$) error bars refer
  to $\Sigma\,m_{\nu} (eV)=0.54,0.4,0.15$, respectively. The black dashed line 
shows the prediction for a massless neutrino cosmology consistent with WMAP-5 
data. Four different
  cases are reported here for Planck+BOSS (top left), Planck + BigBOSS
  (bottom left), ACTPOL+BOSS (top right) and ACTPOL+BigBOSS (bottom right). The redshift evolution of $A$ is here taken into account.}
\end{figure*}

Massive neutrinos are known to suppress the growth of structure in the
early universe on intermediate to small scales $k\gtrsim 10^{-2}$ $\hMpcI$
\cite{Lesgourgues:2006nd}. Since $\langle\delta\flux^2\kappa\rangle$
is mostly sensitive to the same range of scales, it seems reasonable
to examine to what extent massive neutrinos will alter the
$\langle\delta\flux^2\kappa\rangle$ signal. The argument could also be
turned around, asking how well a measurement of
$\langle\delta\flux^2\kappa\rangle$ would allow to constrain the sum
of the neutrino masses. In this first work, we take the first route
and we simply calculate how the $\langle\delta\flux^2\kappa\rangle$
signal is affected by different values of the neutrino masses. We
leave the analysis of the constraining power of
$\langle\delta\flux^2\kappa\rangle$ to a forthcoming work.

Quite generally massive neutrinos affect the matter density power
spectrum in a scale dependent way (see \cite{Lesgourgues:2006nd} for a
review). To account for this effect in an exact way it would require
substantial modifications of the formalism and of the code that we are
currently using to evaluate $\langle\delta\flux^2\kappa\rangle$. In
particular, it would not be possible any longer to separate the
integrations over the comoving distance from the ones over the
wavenumbers $k$. We leave this important development to a future
project and for the purpose of this work we rely on the following
approximation \cite{Hu:1997vi} for the growth of the dark matter
perturbations
\begin{equation}
 \delta_{\textrm{cdm}}\propto D(a)^{1-\frac{3}{5}f_{\nu}},
\end{equation} 
where $f_{\nu}\equiv\Omega_{\nu}/\Omega_m$.  The above expression
should be accurate from the very large scales down to those mildly non-linear 
ones of the \Lya forest. Departures
at small scales are best handled with N-body or hydrodynamical
codes \cite{hanne08}.


The second aspect that we need to take into account before proceeding
with the calculation is that consistency with CMB data requires that a change in 
the sum of the neutrino masses is accompanied by a change in the power 
spectrum normalization $\sigma_8$ \cite{Komatsu:2008hk}. This fact has a 
profound consequence. Just by counting the number of powers of the power 
spectrum that enter in the different expressions, it is straightforward to note that $
\langle\delta\flux^2\kappa\rangle\sim\sigma_8^4$, that $\sigma^2_{\langle\delta
\flux^2\kappa\rangle}\sim\sigma_8^6$ and that its S/N ratio is proportional to $
\sigma_8$. Consequently, a change in the neutrino masses, which requires a 
change in $\sigma_8$ to maintain consistency with CMB data, will cause a 
change in $\langle\delta\flux^2\kappa\rangle$.

To
take this into account we proceed as follows. First we consider the
set of values allowed by the WMAP-5 data in the $\sigma_8-\Sigma
m_{\nu}$ space at 95\% CL. These correspond the the dark red area of
the center panel of Fig. 17 in Komatsu et
al.~\cite{Komatsu:2008hk}. We then choose three flat models with massive
neutrinos consistent with the WMAP-5 data and we use CAMB to generate
the respective dark matter power spectra to be used in the
calculation. The value of the cosmological parameters used for each
model are summarized in Tab.~\ref{Table:nu_cosm_par}. 


\begin{table}
\begin{tabular}{ccccccc}
\hline
Num. & $\Omega_m$ & $\Omega_{\Lambda}$ & $\Omega_{\nu}$ & $\Sigma\,m_
{\nu} (eV)$ & $\sigma_8$ & h \\
\hline\hline
1 & 0.269 & 0.719 & 1.2e-2  & 0.54  & 0.657 & 0.70\\
2 & 0.269 & 0.722 & 8.8e-3  & 0.40  & 0.708 & 0.70\\
3 & 0.269 & 0.728 & 3.3e-3  & 0.15  & 0.786 & 0.70\\
4 & 0.256 & 0.744 &  0.0 & 0.0 & 0.841 & 0.72\\
\hline
\end{tabular} 
\caption{\label{Table:nu_cosm_par} Values of the cosmological
  parameters assumed to estimate the effect of massive neutrinos on
  $\langle\delta\flux^2\kappa\rangle$. All models assume flat geometry.}
\end{table}

One last point is left to be considered. Note in fact that the S/N
ratio for $\langle\delta\flux^2\kappa\rangle$, although increasing
with $\Delta z$, does not increase at a very high rate. It seems
therefore possible to speculate that subdividing the \Lya spectra into
sub-spectra, each of length $dz=0.1$, despite lowering the S/N ratio
for each single sub-spectrum, would allow to reach a better measurement
of the redshift dependence of the signal.

Figure \ref{Fig:Neutrinos} below shows the result of applying the
latter procedure.
The black, orange and red data points represents predicted values of the $\langle
\delta\flux^2\kappa\rangle$ correlator for values of $\sum\,m_{\nu}=
\{0.54,0.4,0.15\}$ respectively, while the dashed black line shows the value of the 
correlator for a $\Lambda$CDM cosmology with massless neutrinos.
As one can see, the cross-correlation signal is quite sensitive to the presence of 
massive
neutrinos and already BOSS and Planck could provide constraints on the
strength of such correlators. As pointed out above, this is due to the fact that more 
massive neutrinos requires smaller values of $\sigma_8$, which in turn 
depresses the signal.

It is here necessary to point out one important caveat. In this paper, we are making a tree-level approximation to the growth rate of $k$ modes: this enables us to separate integrations along the comoving distances from
integrations on the different modes. As previously mentioned, this
approximation does not include the scale-dependent effects of neutrinos on the growth rate of structure. Similarly, this approximation also does not allow us to
take into account the non-linearities induced by gravitational
collapse, which on the other hand tend to enhance the power spectrum
on small scales. We will need to either use  Hyper-Extended Perturbation Theory results or non-linear simulations to evaluate these effects.

\subsection{Early Dark Energy}
\label{ede}

\begin{figure}
\includegraphics[width=0.99\columnwidth]{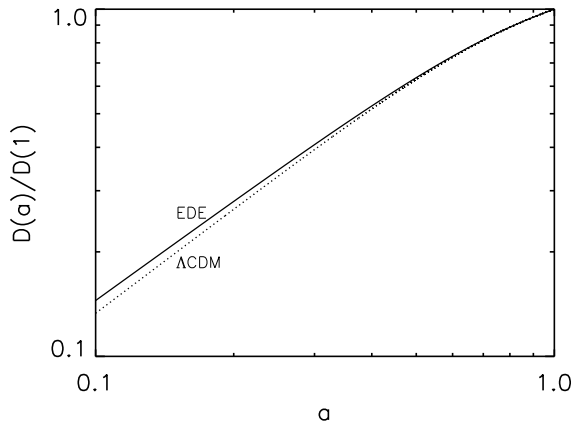}
\caption{\label{Fig:EDE_model} Growth factors for the WMAP-5 flat
  $\Lambda$CDM cosmology (dotted curve) and for the early dark energy
  (EDE) model assumed in this section for comparison (solid curve).}
\end{figure}

Since early dark energy or deviations from general relativity affect the growth rate of structure as a function of scale, the measurements of $\langle
\delta\flux^2\kappa\rangle(z)$ can in principle probe these effects.  Here we focus on early dark energy
(EDE) models, where dark energy makes a significant contribution to the energy density of the universe over a wide range of redshifts. The differences between EDE models and pure $\Lambda$CDM are particularly evident at
high redshifts, when the former has been shown to influence the growth of the first
cosmic structures both in the linear and in the non-linear regime.

We consider here the EDE model proposed in \cite{linder06} and
recently constrained by \cite{Xia:2009ys} (model EDE1 of \cite{Xia:2009ys}). We
compare this model with the $\Lambda$CDM cosmology assumed until
now. The differences in the growth factors for these two
models is shown in Fig.~\ref{Fig:EDE_model} (the difference in the
Hubble parameter evolution is smaller).

\begin{figure*}
\includegraphics[width=0.49\textwidth]{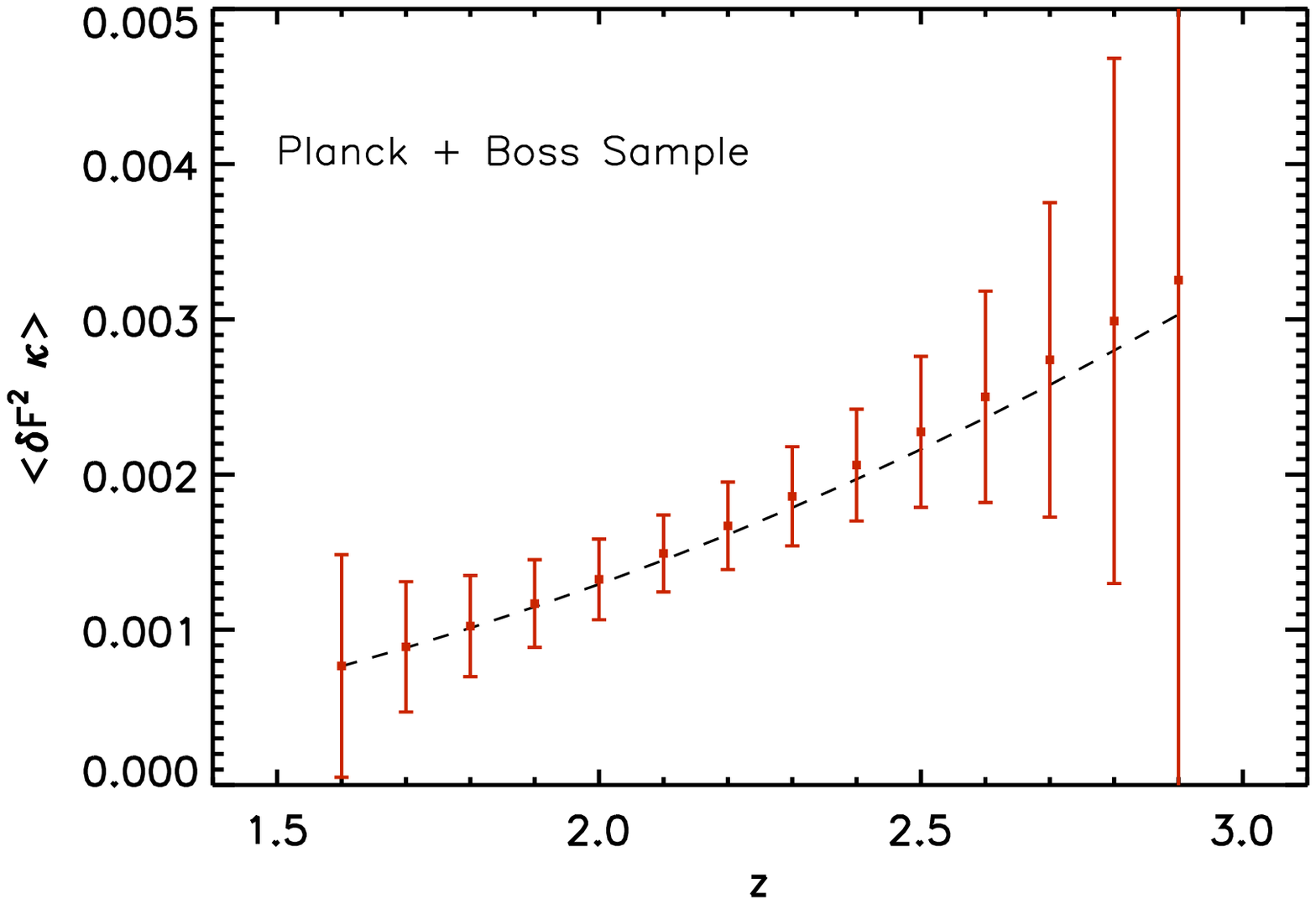}\includegraphics[width=0.49\textwidth]{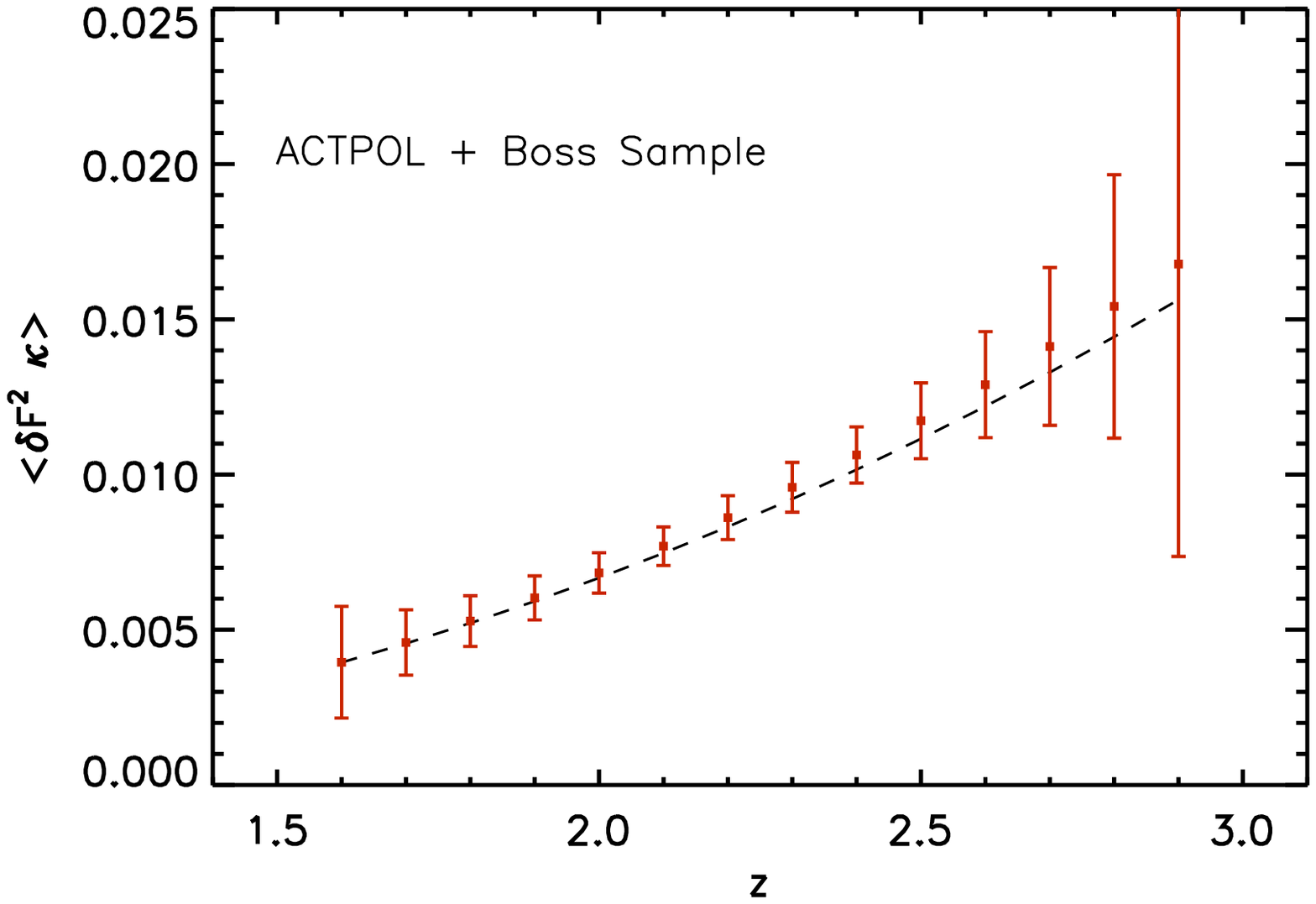}
\includegraphics[width=0.49\textwidth]{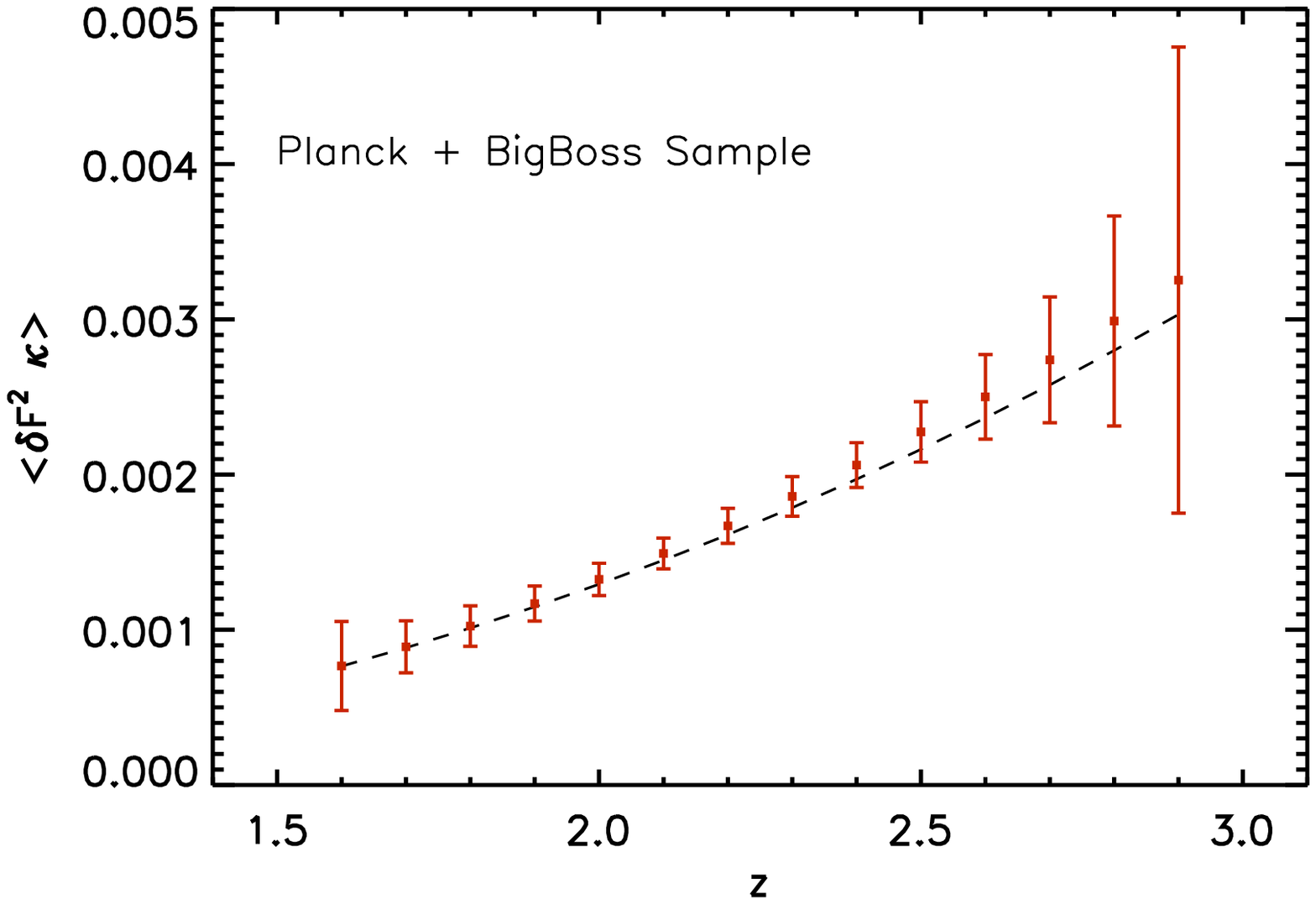}
\includegraphics[width=0.49\textwidth]{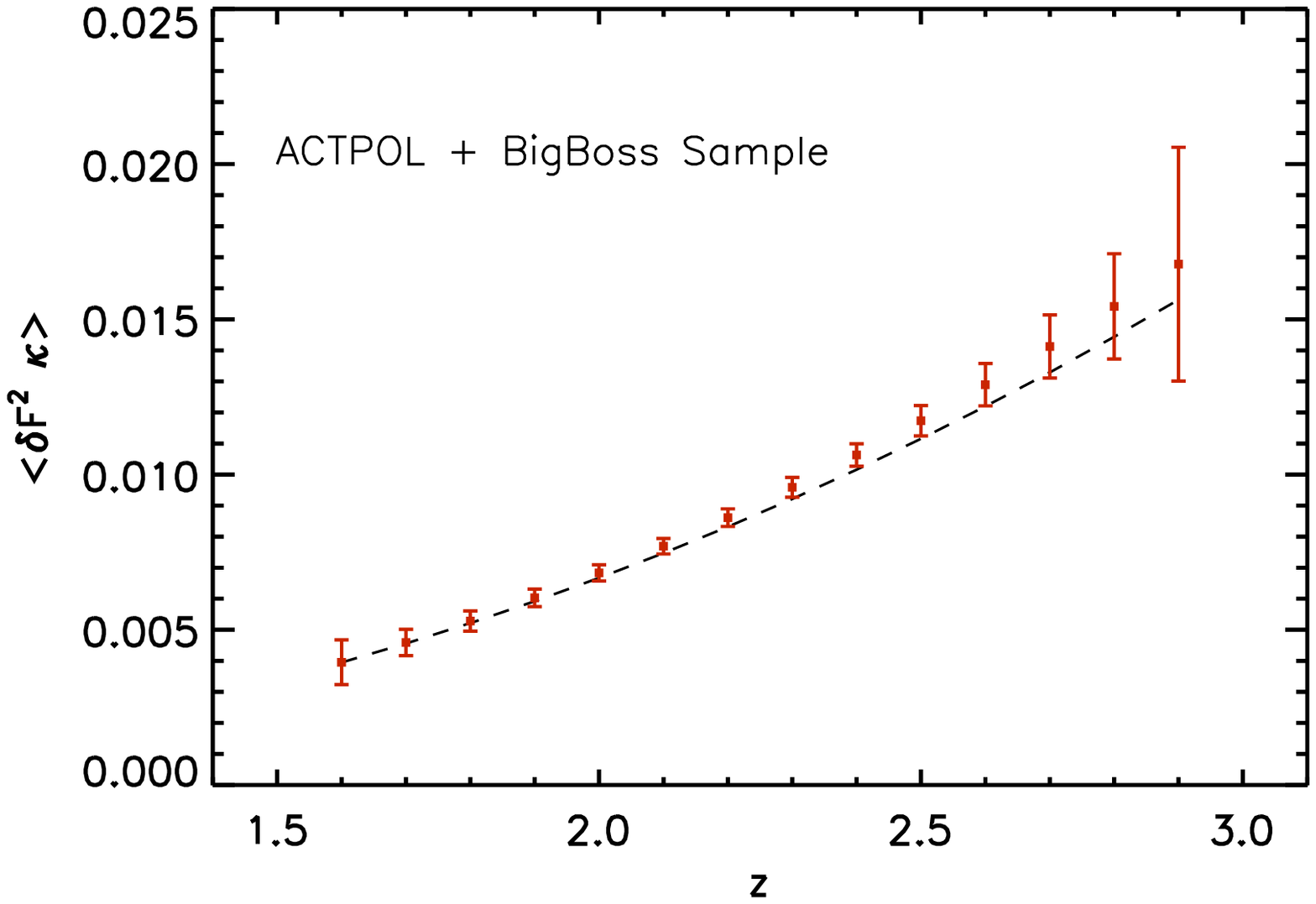}
\caption{\label{Fig:EDE_Pol} Value of $\langle\delta\flux^2\kappa\rangle$ 
estimated for the early
  dark energy EDE model of \cite{linder06}. The dashed black line shows the 
expected value of the correlator for the $\Lambda$CDM cosmology assumed so 
far. Four
  different cases are reported here for Planck+BOSS (top left), Planck
  + BigBOSS (bottom left), ACTPOL+BOSS (top right) and ACTPOL+BigBOSS
  (bottom right). The redshift evolution of $A$ is here taken into account.}
\label{figede}
\end{figure*}

We quantify the departure of the correlators predicted for the EDE
model from the $\Lambda$CDM one using the following expression:
\begin{equation}
 \Delta\chi^2=\sum_i\frac{\left(\langle\delta\flux^n\kappa\rangle_{\rm
     EDE}-\langle\delta\flux^n\kappa\rangle_{\Lambda {\rm CDM}}\right)^2}{\sigma_{{\rm EDE},i}^2} \,.
\end{equation} 

The results are shown in Fig.~\ref{figede} and are summarized in
Table IV. In this case the differences between EDE and $\Lambda$CDM
are very limited and could only be appreciated at some
significance with an advanced CMB experiment like ACTPOL and by increasing the
number of spectroscopic QSOs with BigBOSS. However, it is worth
stressing that the two models presented here are in perfect agreement
with all the low redshift probes and the large-scale structure
measurements provided by galaxy power spectra, CMB, Type Ia supernovae and \Lya forest. Therefore, possible
departures from $\Lambda$CDM can be investigated only exploiting the
capabilities of this intermediate redshift regime with such
correlations or with similar observables in this redshift range.

\begin{table}
\begin{tabular}{ccc}
\hline
QSO sample & CMB Experiment & $\Delta\chi^2$ \\
\hline\hline
$1.6\cdot10^5$ (BOSS) & Planck & 0.3451\\
$1.6\cdot10^5$ (BOSS) & ACTPOL & 2.157\\
$1.0\cdot10^6$ (BigBOSS) & Planck & 1.458\\
$1.0\cdot10^6$ (BigBOSS) & ACTPOL & 9.117\\
\hline
\end{tabular} 
\caption{\label{Table:DChi2} Summary of the estimated $\Delta\chi^2$
  between EDE and $\Lambda$CDM for four different combinations of
  future QSO and CMB experiments using the $\langle\delta\flux^2\kappa\rangle$ 
correlator.}
\end{table}

\section{Conclusions}
\label{discuss}
This work presents a detailed investigation of the cross-correlation signals 
between transmitted \Lya flux and the weak lensing convergence of the CMB along the same 
line-of-sight. One of the motivations behind this work is that the \Lya forest has 
already been shown to be a powerful cosmological tool and novel ways of 
exploring and deepening the understanding of the flux/matter relation could 
significantly improve our knowledge of the high redshift universe. These 
correlators are able to provide astrophysical and cosmological information: since 
they are sensitive to both the flux/matter relation and the value of cosmological 
parameters, in principle they can be used to put constraints on both.

The correlators investigated in the present work have a clear physical meaning. 
The correlation of $\delta\flux$ with $\kappa$ measures to what extent the 
fluctuations along the los mapped by the \Lya forest contribute to the CMB 
convergence field. This correlation is dominated by long wavelength modes ($k
\lesssim 10^{-1}\,\hMpcI$) and as such is more sensitive to \Lya forest continuum 
fitting procedures. The correlation of the flux variance $\delta\flux^2$ with $\kappa
$ measures to what extent the growth of short wavelength modes (mapped by the 
\Lya flux) is enhanced or depressed by the fact that the latter are sitting in regions 
that are overdense or underdense on large scales. This interplay between short 
and long wavelength modes is well exemplified by the redshift dependence of the 
S/N ratio for $\langle\delta\flux^2\kappa\rangle$: lowering the redshift increases 
the S/N ratio because while the variance of $\langle\delta\flux^2\kappa\rangle$ is 
dominated by the independent growth of long and short wavelength modes, the 
value of $\langle\delta\flux^2\kappa\rangle$ itself receives an extra contribution 
due to the fact that the growth of the short wavelength modes is enhanced by the 
presence (and independent growth) of the long wavelength modes. Furthermore, 
this correlator is sensitive to intermediate-to-small scales ($k\gtrsim 10^{-2}\,
\hMpcI$) and as such it should be less sensitive to \Lya forest continuum fitting 
procedures.

To estimate the values of the correlators, their variance and their S/N ratio we 
rely on linear theory and simple approximations, such as the fluctuating Gunn-Peterson approximation at first order. Although the framework is simplified, the 
results are by no means obvious since different modes enter non-trivially in these 
quantities and in their signal-to-noise ratio. We estimate that such correlations 
may be detectable at a high significance level by Planck and the SDSS-III BOSS survey, experiments that 
are already collecting data.   
Moreover, our investigation of the modes of the \Lya forest that contribute to $
\langle\delta\flux^2\kappa\rangle$ shows that the low-resolution \Lya spectra 
measured by SDSS-III (which is aimed at the measurement of BAO at $z=2-4$ 
\cite{mcdonald.eisenstein:2007,slosar09}) should have enough resolution to yield a 
significant S/N.

The peculiar dependence of $\langle\delta\flux^2\kappa\rangle$ on intermediate-
to-short scales and its sensitivity to the value of the power spectrum normalization 
$\sigma_8$ makes it a very useful cosmological tool to test all models 
characterized by variations of the power spectrum on such scales. In particular, 
we applied our estimates to evaluate the sensitivity of $\langle\delta\flux^2\kappa
\rangle$ to changes in $\sigma_8$ due to variations in the sum of the neutrino 
masses and to show how promising this measurement could be in constraining 
the latter. 

Finally, some caveats are in order. First, the code developed to estimate $\langle
\delta\flux^2\kappa\rangle$ and its variance is based on the tree-level 
perturbation theory results reported here. As such, the results shown do not take 
into account nonlinearities induced by gravitational collapse. The extension of the 
analytic results to take into account this aspect is actually quite straightforward, as 
it only requires the implementation of the so-called ``HyperExtended Perturbation 
Theory'' for the bispectrum \cite{Scoccimarro:2000ee}. However, the implementation of such changes in a 
numerical code are less trivial, as the integrations over the power spectrum and 
over the comoving distance cannot be factored any longer. We have nonetheless 
reason to speculate that the nonlinearities induced by gravitational collapse will not 
dramatically change the picture outlined here. At the redshift range spanned by 
the \Lya forest nonlinearities are normally mild and confined to short scales. 
Furthermore, as shown in Sec.~\ref{spectral}, the S/N ratio for $\langle\delta
\flux^2\kappa\rangle$ dominated by modes with $k\gtrsim 10^{-2} \,\hMpcI$, but 
all decades above $10^{-2} \,\hMpcI$ contribute in the same proportion to both the 
signal \textit{and} its variance. It is therefore conceivable to filter out of the \Lya 
spectra the shortest scales, which are the most affected by nonlinearities and still 
be able to retain a non-negligible S/N.

The second caveat pertains the estimate of the correlators' variance. It is in fact 
necessary to point out that to obtain such \textit{estimates} Wick's theorem has 
been applied. Whether the use of Wick's theorem may or may not lead to an accurate 
result when considering the variance of $\langle \delta\flux^{2} \kappa\rangle$ is debatable. 
On one hand it is possible to point out that the largest part of the signal arises at small separations, where 
the value of the correlator is dominated by its connected part. Analogously, it could be possible to argue that the use of Wick's theorem may lead to underestimating the 
correlators' variance. An exact evaluation of the variance of $\langle \delta\flux^{2} \kappa\rangle$, however, requires the exact calculation of a six point function, 
that to the best of our knowledge has never been determined. On the other hand it is also possible to point out that the connected part of $\langle\delta_c\delta_{c'}\delta_q^2\delta_{q'}^2\rangle$ will be significantly non-zero only when the distances between the different points is small. As such, this term will give a non-zero contribution proportional to the length of the \Lya spectrum, which should be subdominant with respect to the ones considered in section \ref{variance}, that are proportional to the distance from the observer all the way to the last scattering surface.

The third caveat pertains the expansion of the expression for the flux, Eq.~(\ref
{eq:FGPA}). Despite the fact that the expansion carried out in Eq.~(\ref{deltaF}) is 
correct on scales larger than about 1 $\hMpc$, we point out here that the flux as 
expressed in Eq.~(\ref{eq:FGPA}) is intrinsically a non-linear function of the 
overdensity field. It is therefore reasonable to wonder whether the non-linearities 
induced by this non-linear mapping would somehow affect the conclusions 
presented here. A simple way to sidestep the present question is to undo the non-linear mapping by defining a new observable $\hat{\flux}=-\ln(\flux)=A(1+\delta_
{\rm IGM})^{\beta}$ and proceed by measuring its correlations.

The best way to assess to what extent the above caveats affect the estimates 
reported in the present work is through numerical simulations, calculating the 
convergence field on a light cone and at the same time measuring \Lya forest 
synthetic spectra and cross-correlating the two. This will be the next step in our 
investigation and the focus of the next publication.

Finally, on the analytical side we still need to address the estimate of the correlators when the 
power spectrum shows evolution in redshift \textit{and} on different scales at the 
same time. As pointed out, $\langle \delta\flux^{2} \kappa\rangle$ is sensitive to 
scales $k\gtrsim 10^{-2} \,\hMpcI$. As such this correlator is an ideal tool to 
test modifications of gravity that show scale dependent growth. At the same time, 
this development would also allow the implementation of the hyperextended 
perturbation theory results and as such to address analytically the impact of 
gravity induced nonlinearities on the value of the correlators.

\textit{Acknowledgements:} We thank S.~Matarrese, F.~Bernardeau,
S.~Dodelson, J.~Frieman, E.~Sefusatti, N.~Gnedin, R.~Scoccimarro, S.~Ho, D.~Weinberg and
J.~P.~Uzan for useful conversations. AV is supported by the DOE  at Fermilab. MV is supported by grants
PD51, ASI-AAE and a PRIN MIUR. DNS and SD are supported by NSF grant AST/0707731 and NASA theory grant NNX08AH30G. DNS thanks the APC (Paris) for its hospitality in spring 2008 when this project was initiated. AV thanks IAP (Paris) for hospitality during different stages of this project.

\onecolumngrid

\appendix

\section{Derivation of perturbative results for $\langle \delta\flux^2\, \kappa\rangle
$}

In this section we derive the expression for $\langle \delta\flux^2\,
\kappa\rangle$ shown in the text,
Eqs.~(\ref{eq:Def_ddd12_ddd23}-\ref{eq:d2d_23final_kLkl}). We move from
Eq.~(\ref{eq:deltaFmK_1}) and need to find an efficient way to
evaluate $\langle\delta^2(\vn,\chi_q)\delta(\vn,\chi_c)\rangle$. We
start by Fourier transforming this cumulant correlator to get
\begin{eqnarray}
 \langle\delta^2_q\delta_c\rangle&=&
\int \frac{d^3\vec{k}_1}{(2\pi)^3} \, \frac{d^3\vec{k}_2}{(2\pi)^3}\,\frac{d^3\vec{k}_3}
{(2\pi)^3}\
e^{i\left[(\vec{k}_1+\vec{k}_2)\cdot\vec{x}_q+\vec{k}_3\cdot\vec{x}_c\right]}\,
W_{\alpha}(k_{1,\parallel})W_{\alpha}(k_{2,\parallel})W_{\kappa}(k_{3,\perp})
\langle\delta(\vec{k}_1)\delta(\vec{k}_2)\delta(\vec{k}_3)\rangle
\nonumber\\
&=&\int \frac{d^3\vec{k}_1}{(2\pi)^3} \, \frac{d^3\vec{k}_2}{(2\pi)^3}\,\frac{d^3\vec{k}
_3}{(2\pi)^3}\
e^{i\left[(\vec{k}_1+\vec{k}_2)\cdot\vec{x}_q+\vec{k}_3\cdot\vec{x}_c\right]}\,
(2\pi)^3\delta_D^3(\vec{k}_1+\vec{k}_2+\vec{k}_3)
W_{\alpha}(k_{1,\parallel})W_{\alpha}(k_{2,\parallel})W_{\kappa}(k_{3,\perp})
B(\vec{k}_1,\vec{k}_2,\vec{k}_3)
\nonumber\\
&=&\int \frac{d^3\vec{k}_1}{(2\pi)^3} \, \frac{d^3\vec{k}_2}{(2\pi)^3}\,\frac{d^3\vec{k}
_3}{(2\pi)^3}\
e^{i\left[(\vec{k}_1+\vec{k}_2)\cdot\vec{x}_q+\vec{k}_3\cdot\vec{x}_c\right]}\,
(2\pi)^3\delta_D^3(\vec{k}_1+\vec{k}_2+\vec{k}_3)
W_{\alpha}(k_{1,\parallel})W_{\alpha}(k_{2,\parallel})W_{\kappa}(k_{3,\perp})\vs
&\times&2\left[\,F_2(\vec{k}_1,\vec{k}_2)P_L(\vec{k}_1,\chi_1)\,P_L(\vec{k}
_2,\chi_2)
+\,F_2(\vec{k}_2,\vec{k}_3)P_L(\vec{k}_2,\chi_2)\,P_L(\vec{k}_3,\chi_3)+F_2(\vec
{k}_3,\vec{k}_1)P_L(\vec{k}_3,\chi_3)\,P_L(\vec{k}_1,\chi_1)\right].\vs
\label{d2d_fullexpression}
\end{eqnarray}
In the second line we introduced the bispectrum $B(\vec{k}_1,\vec{k}_2,\vec{k}
_3)$, while in the third line we replaced the bispectrum with the expression for its 
kernel $F_2$ and products of the linear matter power spectrum $P_L(\vec{k},\chi)
$. For sake of brevity, we keep implicit the dependence of the window functions 
on the cutoff scales: $W_{\alpha}(k_{i,\parallel})=W_{\alpha}(k_{i,
\parallel},k_L,k_l)$ and $W_{\kappa}(\vec{k}_{i,\perp})=W_{\kappa}(\vec{k}_{i,
\perp},k_C)$.
Next, we point out that the evaluation of Eq.~(\ref{d2d_fullexpression}) requires in 
general the integration over a six dimensional $k$-space, which is further 
complicated by the fact that the different window functions break the spherical 
symmetry that one would normally exploit.

In what follows we adopt the tree level approximation to the bispectrum kernel, 
\begin{eqnarray}
F_2(\vec{k}_i,\vec{k}_j)&=&\frac{5}{7}\,
+\frac{1}{2}\frac{\vec{k}_i\cdot\vec{k}_j}{k_i^2\,k_j^2}(k_i^2+k_j^2)\,
+\frac{2}{7}\left(\frac{\vec{k}_i\cdot\vec{k}_j}{k_i\,k_j}\right)^2,\,\label{F2}
\end{eqnarray} 
which can readily be obtained from the more general expression derived by 
Scoccimarro and Couchman \cite{Scoccimarro:2000ee} 
\begin{eqnarray}
F_2^{HEPT}(\vec{k}_i,\vec{k}_j)&=&\frac{5}{7}\,a(n,k_i)\,a(n,k_j)
+\frac{1}{2}\frac{\vec{k}_i\cdot\vec{k}_j}{k_i^2\,k_j^2}(k_i^2+k_j^2)\,b(n,k_i)\,b
(n,k_j)
+\frac{2}{7}\left(\frac{\vec{k}_i\cdot\vec{k}_j}{k_i\,k_j}\right)^2\,c(n,k_i)\,c(n,k_j),
\label{F2_HEPT}
\end{eqnarray} 
setting the three auxiliary functions $a(k)$, $b(k)$ and $c(k)$ that allow to account 
for non-linear growth of structure equal to unity. A generalization of the results 
shown below to take into account the more general formulation of Eq.~(\ref{F2_HEPT}) is 
straightforward to derive.

To proceed further we note that each of the three terms appearing in
the square bracket of Eq.~(\ref{d2d_fullexpression}) depend only on
\textit{two} of the three wavevectors. When moving from the second to
the third line, it is then essential \textit{not} to carry out the
integration over the delta function, because for \textit{each} of
these terms we integrate the Dirac $\delta$ in order to obtain an
expression that depends only on the same wavevectors that appear in
the $F_2$ kernel. The fact that two of the three physical points are the same also
spoils the cyclic symmetry of the bispectrum. In particular, the
$\{1,2\}$ term will differ from the $\{2,3\}$ and $\{3,1\}$ terms. We therefore let
\begin{equation}
 \langle\delta^2_q\delta_c\rangle=\langle\delta_q^2\delta_c\rangle_{1,2}+2\langle
\delta_q^2\delta_c\rangle_{2,3},
\end{equation} 
and start by considering
$\langle\delta^2_q\delta_c\rangle_{1,2}$. Integrating over the
$\delta_D$ function in order to get rid of $\vec{k}_3$ in favor of
$\vec{k}_1$ and $\vec{k}_2$, and then adopting a cylindrical
coordinate system in $k$-space we get
\begin{eqnarray}
 \langle\delta^2\delta\rangle_{1,2}&=&2\,\int
 \frac{dk_{1,\parallel}}{2\pi}\frac{dk_{2,\parallel}}{2\pi}
 e^{i(k_{1,\parallel}+k_{2,\parallel})\Delta\chi\,}W_{\alpha}(k_{1,\parallel})\,
 W_{\alpha}(k_{2,\parallel}) \int_{|k_{1,\parallel}|}^{\infty}
 \frac{k_1 dk_1}{(2\pi)^2}
 P(\vec{k}_1,\chi_1)\,\int_{|k_{2,\parallel}|}^{\infty} \frac{k_2
   dk_2}{(2\pi)^2}P(\vec{k}_2,\chi_2)\,\vs &\times&\int d\phi\,\int
 d\theta_{\perp}
 F_2(\vec{k}_1,\vec{k}_2)\,W_{\kappa}[|\vec{k}_{1,\perp}+\vec{k}_{2,\perp}|].\label
{d2d_window1}
\end{eqnarray} 
As also recognized in \cite{Bernardeau:1995ty}, the most challenging part
of the calculation consists of the integration over the angular variables. This is
because the convergence window function depends on
$|\vec{k}_{1,\perp}+\vec{k}_{2,\perp}|$. The integration over the
angular variables in this case does not necessarily lead to an
expression that may be numerically efficient to evaluate. In
particular we aim to keep integrations factored as much as
possible. Our first goal then is to integrate
\begin{eqnarray}
\int d\phi\,\int d\theta_{\perp}
F_2(\vec{k}_1,\vec{k}_2)\,W_{\kappa}[|\vec{k}_{1,\perp}+\vec{k}_{2,\perp}|]&=&2\pi
\,
\exp\left(-\frac{k^2_{1,\perp}}{k_C^2}\right)\,
\exp\left(-\frac{k^2_{2,\perp}}{k_C^2}\right)\vs &\times& \int
d\theta_{\perp}
F_2(\vec{k}_1,\vec{k}_2)\,\exp\left[-2\frac{k_{1,\perp}k_{2,\perp}\cos(\theta_
{\perp})}{k_C^2}\right],
\end{eqnarray} 
where $\theta_{\perp}$ is the angle between $\vec{k}_{1,\perp}$ and $\vec{k}_
{2,\perp}$.
Now, as far as the integration over the angular variable is concerned, the kernel 
$F_2$ can be written as
\begin{equation}
 F_2(\vec{k}_1,\vec{k}_2)=R+S\,\cos(\theta_{\perp})+T\,\cos^2(\theta_{\perp}),
\end{equation} 
where we have decomposed $\vec{k}$ into its component parallel and
perpendicular to the los according to
$\vec{k}=k_{\parallel}\vn+\vec{k}_{\perp}$ and extracted the terms
that are proportional to different powers of $\cos(\theta_{\perp})$
\begin{eqnarray}
 R&=&\frac{5}{7}\,
+\frac{1}{2}\frac{k_{1,\parallel}\,k_{2,\parallel}}{k_1^2\,k_2^2}(k_1^2+k_2^2)\,
+\frac{2}{7}\left(\frac{k_{1,\parallel}\,k_{2,\parallel}}{k_1\,k_2}\right)^2,\\
 S&=&\frac{1}{2}\frac{k_{1,\perp}\,k_{2,\perp}}{k_1^2\,k_2^2}(k_1^2+k_2^2)+\frac
{4}{7}\frac{k_{1,\parallel}\,k_{2,\parallel}\,k_{1,\perp}\,k_{2,\perp}}{k_1^2\,k_2^2},\\
T&=&\frac{2}{7}\left(\frac{k_{1,\perp}\,k_{2,\perp}}{k_1\,k_2}\right)^2.
\end{eqnarray} 
Integration over the angular variable can then be carried out by remembering that
\begin{eqnarray}
 \int_0^{2\pi}d\theta\,\exp\left[-\alpha\cos(\theta)\right]&=&2\pi\,I_0(\alpha),\\
 \int_0^{2\pi}d\theta\,\exp\left[-\alpha\cos(\theta)\right] \cos(\theta)&=&-2\pi\,I_1
(\alpha),\\
\int_0^{2\pi}d\theta\,\exp\left[-\alpha\cos(\theta)\right] \cos^2(\theta)&=&\frac{2\pi}
{\alpha}\left[\,I_1(\alpha)+\alpha\,I_2(\alpha)\right],
\end{eqnarray} 
where $I_n$ denotes the modified Bessel function of the first kind and
n-th order. The integration over the angular variables yields
\begin{align*}
&\int d\phi\,\int d\theta_{\perp}
F_2(\vec{k}_1,\vec{k}_2)\,W_C[|\vec{k}_{1,\perp}+\vec{k}_{2,\perp}|]
= (2\pi)^2\, \exp\left(-\frac{k^2_{1,\perp}}{k_C^2}\right)\,
\exp\left(-\frac{k^2_{2,\perp}}{k_C^2}\right)\vs
&\times \left\{R\,I_0\left(2\frac{k_{1,\perp}k_{2,\perp}}{k_C^2}\right)
-S\,I_1\left(2\frac{k_{1,\perp}k_{2,\perp}}{k_C^2}\right)+T\left[\frac{k_C^2}{2\,k_
{1,\perp}k_{2,\perp}}I_1\left(2\frac{k_{1,\perp}k_{2,\perp}}{k_C^2}\right)+I_2\left
(2\frac{k_{1,\perp}k_{2,\perp}}{k_C^2}\right)\right]\right\}. 
\end{align*} 
The difficulty with this result is that every term depend on the
product $k_{1,\perp} k_{2,\perp}$. As such, we are facing a 2D
\textit{joint} integration over the whole $[k_{1,\perp},k_{2,\perp}]$
domain. If on one hand this is doable, on the other hand we are more
interested in obtaining a final result which is a product of integrals
instead of the integral of the product. It is possible to move around
this obstacle recalling that (Abramowitz and Stegun \cite{Abramowitz:1965hc}, 
9.6.10)
\begin{eqnarray}
 I_{\nu}(z)&=&\sum_{n=0}^{\infty}\frac{1}{n!\Gamma(\nu+n+1)}\left(\frac{z}{2}\right)^
{2n+\nu}=\sum_{n=0}^{\infty}I_{\nu}^{(n)}\left(\frac{z}{2}\right)^{2n+\nu}.
\end{eqnarray} 
We can write the modified Bessel function splitting the dependence on $k_
{1,\perp}$ and $k_{2,\perp}$ as
\begin{eqnarray}
 I_0\left(2\frac{k_{1,\perp}k_{2,\perp}}{k_C^2}\right)&=&\sum_{n=0}^{\infty}I_{0}^
{(n)}\left(\frac{k_{1,\perp}^2}{k_C^2}\right)^{n}\left(\frac{k_{2,\perp}^2}
{k_C^2}\right)^{n},\\
 I_1\left(2\frac{k_{1,\perp}k_{2,\perp}}{k_C^2}\right)&=&\frac{k_{1,\perp}k_
{2,\perp}}{k_C^2}\sum_{n=0}^{\infty}I_{1}^{(n)}\left(\frac{k_{1,\perp}^2}
{k_C^2}\right)^{n}\left(\frac{k_{2,\perp}^2}{k_C^2}\right)^{n},\\
I_2\left(2\frac{k_{1,\perp}k_{2,\perp}}{k_C^2}\right)&=&\left(\frac{k_{1,\perp}k_
{2,\perp}}{k_C^2}\right)^2\sum_{n=0}^{\infty}I_{2}^{(n)}\left(\frac{k_{1,\perp}^2}
{k_C^2}\right)^{n}\left(\frac{k_{2,\perp}^2}{k_C^2}\right)^{n},
\end{eqnarray}
where for sake of brevity we use the following notation for the coefficients
\begin{eqnarray}
 I_0^{(n)}&=&\frac{1}{n!^2},\\
 I_1^{(n)}&=&\frac{1}{n!(n+1)!}=\frac{I_0^{(n)}}{n+1},\\
 I_2^{(n)}&=&\frac{1}{n!(n+2)!}=\frac{I_0^{(n)}}{(n+1)(n+2)}.
\end{eqnarray} 
Now, however complicated, this form allows us to factor the different
integrals. Let's start by considering the term $R\,I_0$. We have
\begin{eqnarray}
 R\,I_0&=&\sum_{m=0}^{\infty}I_{0}^{(m)}\left(\frac{k_{1,\perp}^2}{k_C^2}\right)^{m}
\left(\frac{k_{2,\perp}^2}{k_C^2}\right)^{m}
\left[
\frac{5}{7}
+\frac{1}{2}\frac{k_{1,\parallel}\,k_{2,\parallel}}{k_1^2\,k_2^2}(k_1^2+k_2^2)
+\frac{2}{7}\left(\frac{k_{1,\parallel}\,k_{2,\parallel}}{k_1\,k_2}\right)^2
\right]\nonumber\\
&=&\sum_{m=0}^{\infty}I_{0}^{(m)}\left(\frac{k_1^2-k_{1,\parallel}^2}{k_C^2}\right)^
{m}\left(\frac{k_2^2-k_{2,\parallel}^2}{k_C^2}\right)^{m}\frac{5}{7}\nonumber\\
&+&\sum_{m=0}^{\infty}I_{0}^{(m)}\left(\frac{k_1^2-k_{1,\parallel}^2}{k_C^2}\right)^
{m}\left(\frac{k_2^2-k_{2,\parallel}^2}{k_C^2}\right)^{m}
\left[\frac{1}{2}\frac{k_{1,\parallel}\,k_{2,\parallel}}{k_1^2\,k_2^2}
(k_1^2+k_2^2)\right]\nonumber\\
&+&\sum_{m=0}^{\infty}I_{0}^{(m)}\left(\frac{k_1^2-k_{1,\parallel}^2}{k_C^2}\right)^
{m}\left(\frac{k_2^2-k_{2,\parallel}^2}{k_C^2}\right)^{m}
\left[\frac{2}{7}\left(\frac{k_{1,\parallel}\,k_{2,\parallel}}{k_1\,k_2}\right)^2\right],
\end{eqnarray} 
where in going from the first to the second step we expressed
$k_{\perp}^2$ as a function of $k$ and $k_{\parallel}$ using the fact
that $k^2=k_{\parallel}^2+k_{\perp}^2$. This is necessary because the
power spectrum is function of $k$ and not of $k_{\perp}$.  We can then
proceed by defining the following functions
\begin{eqnarray}
\tilde{H}_m(k_{\parallel},\chi;k_C)&\equiv&\int_{|k_{\parallel}|}^{\infty}\frac{k\,dk}
{2\pi}\,\sqrt{I_0^{(m)}}\,P(k,\chi)\,\left(\frac{k^2-k_{\parallel}^2}{k_C^2}\right)^{m}\,
\exp\left(-\frac{k^2-k_{\parallel}^2}{k_C^2}\right),\label{def:Htilde}\\
\tilde{L}_m(k_{\parallel},\chi;k_C)&\equiv&
\int_{|k_{\parallel}|}^{\infty}\frac{dk}{2\pi k} \,\sqrt{I_0^{(m)}}\,P(k,\chi)\,\left(\frac{k^2-
k_{\parallel}^2}{k_C^2}\right)^{m}\,\exp\left(-\frac{k^2-k_{\parallel}^2}
{k_C^2}\right).\label{def:Ltilde}
\end{eqnarray}
It is important to notice that because of the integration domain \textit{all} the above 
functions are \textit{even} in $k_{\parallel}$, regardless of the value of $m$.
With the help of these functions we then have
\begin{eqnarray}
&& \int_{|k_{1,\parallel}|}^{\infty} 
\frac{k_1 dk_1}{(2\pi)^2} P(\vec{k}_1,\chi_1)\,\int_{|k_{2,\parallel}|}^{\infty} \frac{k_2 
dk_2}{(2\pi)^2}P(\vec{k}_2,\chi_2)
\int d\phi\,\int d\theta_{\perp}
R\,W_{\kappa}[|\vec{k}_{1,\perp}+\vec{k}_{2,\perp}|]\nonumber\\
&=&\frac{5}{7}\sum_{m=0}^{\infty}\tilde{H}_m(k_{1,\parallel},\chi_1)\tilde{H}_m(k_
{2,\parallel},\chi_2)
+\frac{2\,k_{1,\parallel}^2\,k_{2,\parallel}^2}{7}\sum_{m=0}^{\infty}\tilde{L}_m(k_
{1,\parallel},\chi_1)\tilde{L}_m(k_{2,\parallel},\chi_2)\nonumber\\
&+&\frac{k_{1,\parallel}\,k_{2,\parallel}}{2}\left[\sum_{m=0}^{\infty}
\tilde{H}_m(k_{1,\parallel},\chi_1)\tilde{L}_m(k_{2,\parallel},\chi_2)
+\sum_{m=0}^{\infty}\tilde{L}_m(k_{1,\parallel},\chi_1)\tilde{H}_m(k_{2,\parallel},
\chi_2)\right].
\end{eqnarray}
We have therefore succeeded in obtaining an expression that has the
dependence on $k_{1,\parallel}$ and $k_{2,\parallel}$
\textit{completely factored}. The sums over $m$ and the fact that each
term is a product of factors that only depend either on
$k_{1,\parallel}$ or on $k_{2,\parallel}$ allows an integration term
by term and at the same time to bypass the two dimensional joint
integration.

We can then proceed exactly in the same way for the other two terms,
$S\,I_1$ and $T\,I_2$ with the only difference that in order to obtain
expressions where only the coefficients of the modified Bessel
function of $0$-th order $I_0^{(m)}$ appear we use the fact that
$I_1^{(m)}=(m+1)\,I_0^{(m+1)}$. We then obtain for the $S$ term the
following expression
\begin{eqnarray}
&& \int_{|k_{1,\parallel}|}^{\infty} 
\frac{k_1 dk_1}{(2\pi)^2} P(\vec{k}_1,\chi_1)\,\int_{|k_{2,\parallel}|}^{\infty} \frac{k_2 
dk_2}{(2\pi)^2}P(\vec{k}_2,\chi_2)\,
\int d\phi\,\int d\theta_{\perp}
S\,\cos(\theta_{\perp})\,W_{\kappa}[|\vec{k}_{1,\perp}+\vec{k}_{2,\perp}|]\\
&=&-\frac{k_C^2}{2}\sum_{m=0}^{\infty}m\,\left[\tilde{H}_m(k_{1,\parallel},
\chi_1)\tilde{L}_m(k_{2,\parallel},\chi_2)
+\tilde{L}_m(k_{1,\parallel},\chi_1)\tilde{H}_m(k_{2,\parallel},\chi_2)\right]
-\frac{4\,k_C^2\,k_{1,\parallel}\,k_{2,\parallel}}{7}\sum_{m=0}^{\infty}\,m\,\tilde{L}
_m(k_{1,\parallel},\chi_1)\tilde{L}_m(k_{2,\parallel},\chi_2).\nonumber
\end{eqnarray} 
Finally, the $T$ term gives
\begin{eqnarray}
 &&\int_{|k_{1,\parallel}|}^{\infty} 
\frac{k_1 dk_1}{(2\pi)^2} P(\vec{k}_1,\chi_1)\,\int_{|k_{2,\parallel}|}^{\infty} \frac{k_2 
dk_2}{(2\pi)^2}P(\vec{k}_2,\chi_2)
\int d\phi\,\int d\theta_{\perp}
T\,\cos^2(\theta_{\perp})\,W_C[|\vec{k}_{1,\perp}+\vec{k}_{2,\perp}|]
\nonumber\\
&=&\frac{k_C^4}{7}\sum_{m=0}^{\infty}\,m\,(2m-1)\,\tilde{L}_m(k_{1,\parallel},
\chi_1)\tilde{L}_m(k_{2,\parallel},\chi_2).
\end{eqnarray} 
With the introduction of the definitions (\ref{def:Htilde}-\ref{def:Ltilde}) and with the 
series expansion for the modified Bessel function we have therefore managed to 
carry out the integration over the perpendicular part of the wavevector. We are 
then left with the integration over $k_{\parallel}$. First recall that the window 
functions acting on the \Lya flux are
\begin{eqnarray}
W_{\alpha}(k_{\parallel},k_L,k_l)&\equiv&\left[1-e^{-(k_{\parallel}/k_l)^2}\right]e^{-
(k_{\parallel}/k_L)^2}
=e^{-(k_{\parallel}/k_L)^2}-e^{-(k_{\parallel}/\bar{k})^2},
\end{eqnarray}
and that in Eq.~(\ref{d2d_window1}) they decouple from one another. 
We can proceed further by defining the following function
\begin{eqnarray}
 f^{(n)}_m(\Delta\chi,\chi;k_C,k_L)&\equiv&\int_{-\infty}^{\infty}\frac{dk_{\parallel}}
{2\pi}\,\left(\frac{k_{\parallel}}{k_L}\right)^n\exp\left[-\frac{k_{\parallel}^2}
{k_L^2}+ik_{\parallel}\Delta\chi\right]
\tilde{f}_m(k_{\parallel},\chi;k_C),\label{f_m^n}
\end{eqnarray}
where $f=\{H,L\}$. 
It is straightforward to note that because all the 
tilde functions are even in $k_{\parallel}$, depending on the value of $n$ the
above Fourier transforms are either purely real (if $n$ is even) or purely
imaginary (if $n$ is odd). Furthermore, if $n$ is even the above functions are
\textit{real} and \textit{even}, while if $n$ is odd the above functions are
\textit{imaginary} and \textit{odd}. Carrying out the integration on $k_{\parallel}$ is 
then straightforward, as it just corresponds the replacement $k_{\parallel}^n \tilde
{f}_m(k_{\parallel},\chi;k_C)\rightarrow k_L^n f_m^{(n)}(\Delta\chi,\chi;k_C,k_L)-
\bar{k}^n f_m^{(n)}(\Delta\chi,\chi;k_C,\bar{k})$. Finally, from a computational point 
of view
this approach is rather efficient, as the tilded function need to be calculated
only once and then used to construct the two-index functions.

With the help of these auxiliary functions we can finally \
obtain the following
expression for the cumulant correlator $\langle\delta^2\delta\rangle_{1,2}$
\begin{eqnarray}
 \langle\delta^2\delta\rangle_{1,2}&=&
2\sum_{m=0}^{\infty}\left\{\frac{5}{7}\,
\left[H_m^{(0)}(\Delta\chi,\chi_q;k_C,k_L)-H_m^{(0)}(\Delta\chi,\chi_q;k_C,\bar{k})
\right]^2\right.\nonumber\\
&+&\left[k_L\,H_m^{(1)}(\Delta\chi,\chi_q;k_C,k_L)-\bar{k}\,H_m^{(1)}(\Delta\chi,
\chi_q;k_C,\bar{k})\right]
\left[k_L\,L_m^{(1)}(\Delta\chi, \chi_q;k_C,k_L)-\bar{k}L_m^{(1)}(\Delta\chi, 
\chi_q;k_C,\bar{k})\right]
\vs
&-&m\,k_C^2\,\left[H_m^{(0)}(\Delta\chi,\chi_q;k_C,k_L)-H_m^{(0)}(\Delta\chi,
\chi_q;k_C,\bar{k})\right]\left[L_m^{(0)}(\Delta\chi,\chi_q;k_C,k_L)-L_m^{(0)}(\Delta
\chi,\chi_q;k_C,\bar{k})\right]
\vs
&+&
\frac{2}{7}\left[k_L^2\,L_m^{(2)}(\Delta\chi,\chi_q;k_C,k_L)-\bar{k}^2\,L_m^{(2)}
(\Delta\chi,\chi_q;k_C,\bar{k})\right]^2\vs
&-&\frac{4m}{7}\,
k_C^2\,\left[k_L\,L_m^{(1)}(\Delta\chi,\chi_q;k_C,k_L)-\bar{k}\,L_m^{(1)}(\Delta\chi,
\chi_q;k_C,\bar{k})\right]^2
\vs
&+&\left.\frac{m(2m-1)}{
7}\,k_C^4\,\left[L_m^{(0)}(\Delta\chi,\chi_q;k_C,k_L)-L_m^{(0)}(\Delta\chi,
\chi_q;k_C,\bar{k})\right]^2
\right\}.\label{eq:d2d_12_kLkl_final}
\end{eqnarray}
A cautionary note is in order. As mentioned above, 
the functions defined through Eq.~(\ref{f_m^n}) are purely imaginary if the
index $(n)$ is odd. However, notice that in Eq.~(\ref{eq:d2d_12_kLkl_final}) above 
there
are always two such functions that appear together (as in the case with
$H_m^{(1)} L_m^{(1)}$), thus ensuring that
$\langle\delta^2\delta\rangle_{1,2}$ is always real valued. 

Let's now move to calculate $\langle\delta^2\delta\rangle_{2,3}$. 
Notice incidentally that this term is exactly equal to
$\langle\delta^2\delta\rangle_{3,1}$. We start from the now usual expression
\begin{eqnarray}
 \langle\delta^2\delta\rangle_{2,3}&=&2\,\int \frac{dk_{2,\parallel}}{2\pi}\frac{dk_
{3,\parallel}}{2\pi}
e^{-i\,k_{3,\parallel}\Delta\chi}\,
W_{\alpha}(-k_{2,\parallel}-k_{3,\parallel
})\,
W_{\alpha}(k_{2,\parallel})
\int_{|k_{2,\parallel}|}^{\infty} 
\frac{k_2 dk_2}{(2\pi)^2} P(\vec{k}_2,\chi_2)\,\int_{|k_{3,\parallel}|}^{\infty}
\frac{k_3 dk_3}{(2\pi)^2}P(\vec{k}_3,\chi_3)\,
\vs
&\times&\int d\phi\,\int
d\theta_{\perp}
F_2(\vec{k}_2,\vec{k}_3)\,W_{\kappa}(\vec{k}_{3,\perp}),\label{d2d_window23}
\end{eqnarray} 
where, as previously, we have traded the integrations over 
$k_{i,\perp}$ for the ones over $k_i$. In this case the integration over the
angular variables doesn't pose any problem as the window function $W_{\kappa}
$
is actually a function of $k_{3,\perp}$ only and it can be safely pulled out of the 
angular integrals
\begin{eqnarray}
 \int d\phi \int d\theta_{\perp}
F_2(\vec{k}_2,\vec{k}_3)
&=&\frac{(2\pi)^2}{7}\,\left(5+1\right)
+\frac{(2\pi)^2}{2}\,k_{2,\parallel}\,k_{3,\parallel}
\left(\frac{1}{k_2^2}+\frac{1}{k_3^2}\right)
+\frac{(2 \pi)^2}{7}\,
\left[3\frac{k_{2,\parallel}^2\,k_{3,\parallel}^2}{k_2^2\,k_3^2}-\left(\frac{k_
{2,\parallel}^2}{k_2^2}+\frac{k_{3,\parallel}^2}{k_3^2}\right)\right].\label
{Int_ang_vars_nowindow}
\end{eqnarray} 
It is here necessary to point out that since $W_{\kappa}$ depends
only on $k_{3,\perp}$, the tilded functions that will appear when the
integration over $k_2$ is carried out will contain no filter
function. We characterize this functions by substituting to $k_C$ the
$\infty$ symbol, as to all extent the gaussian filter with
$k_C\rightarrow\infty$ just yields unity. We then have
\begin{eqnarray}
&& \int_{|k_{2,\parallel}|}^{\infty} 
\frac{k_2 dk_2}{(2\pi)^2} P(\vec{k}_2,\chi_2)\,\int_{|k_{3,\parallel}|}^{\infty}
\frac{k_3 dk_3}{(2\pi)^2}P(\vec{k}_3,\chi_3)
\int d\phi\,\int
d\theta_{\perp}
F_2(\vec{k}_2,\vec{k}_3)\,W_{\kappa}(\vec{k}_{3,\perp})\nonumber\\
&=&\int_{|k_{2,\parallel}|}^{\infty} 
\frac{k_2 dk_2}{(2\pi)^2} P(\vec{k}_2,\chi_2)
\int_{|k_{3,\parallel}|}^{\infty} \frac{k_3
dk_3}{(2\pi)^2}P(\vec{k}_3,\chi_3)\exp\left(-\frac{k_{3}^2-k_{3,\parallel}^2}{
k_C^2}\right)\nonumber\\
&\times&(2\pi)^2\left[\frac{5}{7}
+\frac{1}{2}\,k_{2,\parallel}\,k_{3,\parallel}
\left(\frac{1}{k_2^2}+\frac{1}{k_3^2}\right)+\frac{1}{7}
\,\frac{1}{k_2^2\,k_3^2}
\left(2\,k_{2,\parallel}^2\,k_{3,\parallel}^2+k_{2,\perp}^2\,k_{3,\perp}
^2\right)\right]\nonumber\\
&=&\frac{6}{7}\,\tilde{H}_0(k_{2,\parallel},\chi_2;\infty)\tilde{H}_0(k_{3,
\parallel},\chi_3;k_C)
+\frac{1}{2}\,k_{2,\parallel}\,k_{3,\parallel}\tilde{L}_0(k_{2,\parallel},
\chi_2;\infty)\tilde{H}_0(k_{3,\parallel},\chi_3;k_C)
\vs&+&\frac{1}{2}\,k_{2,\parallel}\,k_{3,\parallel}\tilde{H}_0(k_{2,\parallel},
\chi_2;\infty)\tilde{L}_0(k_{3,\parallel},\chi_3;k_C)
+\frac{3}{7}
\,k_{2,\parallel}^2\,k_{3,\parallel}^2\tilde{L}_0(k_{2,\parallel},
\chi_2;\infty)\tilde{L}_0(k_{3,\parallel},\chi_3;k_C)\vs
&-&\frac{1}{7}
k_{3,\parallel}^2\,\tilde{H}_0(k_{2,\parallel},\chi_2;\infty)\,\tilde{L}
_0(k_{3,\parallel},\chi_3;k_C)
-\frac{1}{7}k_{2,\parallel}^2\,\tilde{
L}_0(k_{2,\parallel},
\chi_2;\infty)\tilde{H}_0(k_{3,\parallel},\chi_3;k_C).
\end{eqnarray}

The expression for the window function acting on the \Lya flux is in this case
\begin{eqnarray}
W_{\alpha}(-k_{2,\parallel} -k_{3,\parallel})\,W_{\alpha}(k_{2,\parallel})
&=&\left[1-e^{-\left(\frac{k_{2,\parallel}+k_{3,\parallel}}{k_l}\right)^2}\right]e^{-\left
(\frac{k_{2,\parallel}+k_{3,\parallel}}{k_L}\right)^2}
\left[1-e^{-\left(\frac{k_{2,\parallel}}{k_l}\right)^2}\right]e^{-\left(\frac{k_{2,\parallel}}
{k_L}\right)^2}
\vs
&=&e^{-k_{3,\parallel}^2/k_L^2}\left(e^{-2k_{2,\parallel}^2/k_L^2}-e^{-k_
{2,\parallel}^2/\hat{k}^2}\right)
\sum_n\frac{(-2)^n}{n!}\left(\frac{k_{2,\parallel}}{k_L}\right)^n
\left(\frac{k_{3,\parallel}}{k_L}\right)^n\vs
&+&e^{-k_{3,\parallel}^2/\bar{k}^2}
\left(e^{-2k_{2,\parallel}^2/\bar{k}^2}-e^{-k_{2,\parallel}^2/\hat{k}^2}\right)
\sum_n\frac{(-2)^n}{n!}\left(\frac{k_{2,\parallel}}{\bar{k}}\right)^n
\left(\frac{k_{3,\parallel}}{\bar{k}}\right)^n,
\end{eqnarray}
where we have recast the window function in a combination that is suitable for 
furthering the calculation. Notice in fact that the first and second term in the sum 
differ only by the presence of $k_L$ or $\bar{k}$ in the denominators of the 
exponentials. Furthermore, the terms in square brackets are functions of $k_
{2,\parallel}$ only. We then define the coefficients
\begin{eqnarray}
 \bar{f}_m^{(n)}(\chi;k_L)&\equiv&\int_{-\infty}^{\infty}\frac{dk_{\parallel}}{2\pi}\left
(\frac{k_{\parallel}}{k_L}\right)^n\left[e^{-2k_{\parallel}^2/k_L^2}-e^{-k_{\parallel}
^2/\hat{k}^2}\right]\tilde{f}_m(k_{\parallel},\chi,\infty),\\
\bar{f}_m^{(n)}(\chi;\bar{k})&\equiv&\int_{-\infty}^{\infty}\frac{dk_{\parallel}}{2\pi}\left
(\frac{k_{\parallel}}{\bar{k}}\right)^n\left[e^{-2k_{\parallel}^2/\bar{k}^2}-e^{-k_
{\parallel}^2/\hat{k}^2}\right]
\tilde{f}_m(k_{\parallel},\chi,\infty).
\end{eqnarray} 
A point worth making is that the second expression can be obtained from the first 
one with the substitution $k_L\rightarrow \bar{k}$ in the denominators but \textit
{not} in the expression for $\hat{k}$, hence the necessity of two separate 
definitions.
Considering then the following generic term, it is possible to show that
\begin{eqnarray}
&& \int\frac{dk_2}{2\pi}\frac{dk_3}{2\pi}\,k_2^p\,k_3^q\,
\tilde{f}_i(k_2,\chi_2;\infty)\tilde{g}_j(k_3,\chi_3;k_C)
W_{\alpha}(-k_2 -k_3)\,W_{\alpha}(k_2)\,e^{-ik_3\Delta\chi}\vs
&=&\sum_m\frac{(-2)^m}{m!}
\left[k_L^{(p+q)}g_j^{(q+m)}(\Delta\chi,\chi;k_C,k_L)\bar{f}_i^{(p+m)}(\chi_2,k_L)
+\bar{k}^{(p+q)}g_j^{(q+m)}(\Delta\chi,\chi;k_C,\bar{k})\bar{f}_i^{(p+m)}(\chi_2,\bar
{k})\right],
\end{eqnarray} 
which then leads directly to
\begin{eqnarray}
\langle\delta_q^2\delta_c\rangle_{2,3}&=&
2\sum_{m=0}^{\infty}\frac{(-1)^m\,2^m}{m!}
\left[\frac{6}{7}\bar{H}_0^{(m)}(k_L)H_0^{(m)}(\Delta\chi;k_C,k_L)\right.
+\frac{1}{2}k_L^2\bar{L}_0^{(m+1)}(k_L)\,H_0^{(m+1)}(\Delta\chi;k_C,k_L)\vs
&+&\frac{1}{2}k_L^2\bar{H}_0^{(m+1)}(k_L)\,L_0^{(m+1)}(\Delta\chi;k_C,k_L)
+\frac{3}{7}k_L^4\,\bar{L}_0^{(m+2)}(k_L)\,L_0^{(m+2)}(\Delta\chi;k_C,k_L)\vs
&-&\left.
\frac{k_L^2}{7}
\bar{H}_0^{(m)}(k_L)\,L_0^{(m+2)}(\Delta\chi;k_C,k_L)
-\frac{k_L^2}{7}
\bar{L}_0^{(m+2)}(k_L)\,H_0^{(m)}(\Delta\chi;k_C,k_L)
+ (k_L\rightarrow\bar{k})\right].\label{d2d_23final_kLkl}
\end{eqnarray}
Note that in the above expression while $\bar{f}_m^{(n)}$ are always
real, the $f_m^{(n)}$ can be real or imaginary depending on whether
$n$ is even or odd. However, the fact that $\bar{f}_m^{(n)}$ is zero
whenever the upper index is odd guarantees that
$\langle\delta^2\delta\rangle_{2,3}$ is always real valued. Also,
notice that while the coefficients $\bar{f}_m^{(n)}(\chi_Q;k_L)$ are
decreasing with $m$, the coefficients
$\bar{f}_m^{(n)}(\chi_Q;\bar{k})$ are actually \textit{increasing}
with $m$. However, the $m!$ factor present in the denominator more
than compensates for these increasing coefficients and allows to
truncate the series in an actual calculation.

Finally, it is worth pointing out that the case without cutoff on the long wavelength 
mode is recovered from the above expression simply by setting $k_l=0$ and then 
noticing that in this case $\bar{k}=0$ and that therefore the corresponding terms 
appearing in Eqs.~(\ref{eq:d2d_12_kLkl_final}, \ref{d2d_23final_kLkl}) disappear.

\twocolumngrid

\bibliography{VDSV,cmblensing,projects_new}

\end{document}